\documentclass[aps,nofootinbib,prd,showpacs,class-pre]{revtex4}

\usepackage{graphicx, epsfig}
\usepackage{color}
\usepackage{cancel}
\usepackage{mathrsfs}
\usepackage {amssymb}
\usepackage {amsmath}
\usepackage{float}

\newcommand{\be}{\begin{equation}}
\newcommand{\ee}{\end{equation}}
\newcommand{\br}{\mathbf{r}}
\newcommand{\bw}{\mathbf{w}}
\newcommand{\bd}{\mathbf{d}}
\newcommand{\bx}{\mathbf{x}}
\newcommand{\bX}{\mathbf{X}}
\newcommand{\by}{\mathbf{y}}
\newcommand{\bY}{\mathbf{Y}}
\newcommand{\bp}{\mathbf{p}}
\newcommand{\bb}{\mathbf{b}}
\newcommand{\ba}{\mathbf{a}}
\newcommand{\bea}{\begin{eqnarray}}
\newcommand{\eea}{\end{eqnarray}}

\begin{document}

\begin{flushleft}
KCL-PH-TH/2014-48
\end{flushleft}
\title{Zipping and Unzipping in String Networks: Dynamics of Y-junctions}

\author{Anastasios Avgoustidis$^1$\footnote{anastasios.avgoustidis@nottingham.ac.uk},
  Alkistis Pourtsidou$^{2,3}$\footnote{alkistis.pourtsidou@port.ac.uk},
  Mairi Sakellariadou$^4$\footnote{mairi.sakellariadou@kcl.ac.uk}}
  \affiliation{$^1$School of Physics and Astronomy, University of
  Nottingham, University Park, Nottingham NG7 2RD, UK}
  \affiliation{$^2$Institute of Cosmology \& Gravitation, University of Portsmouth, Burnaby Road, Portsmouth, PO1 3FX, United Kingdom}  
  \affiliation{$^3$Dipartimento di Fisica e Astronomia, Universit\`a di
  Bologna, viale B. Pichat 6/2, I-40127 Bologna, Italy}
  \affiliation{$^4$Department of
  Physics, King's College London, University of London, Strand WC2R 2LS,
  London, UK}

\begin{abstract}
We study, within the Nambu-Goto approximation, the stability of
massive string junctions under the influence of the tensions of three
strings joining together in a Y-type configuration.  The relative
angle $\beta$ between the strings at the junction is in general time-dependent
and its evolution can lead to zipping or unzipping of the three-string
configuration. 
We find that these configurations are stable under
deformations of the tension balance condition at the junction. The 
angle $\beta$ relaxes at its equilibrium value and the junction grows
relativistically.  We then discuss other potential ``unzipping agents" including 
monopole/string forces for long strings and curvature for loops, and we 
investigate specific solutions exhibiting decelerated zipping and unzipping 
of the Y-junction. These results provide motivation for incorporating the effects 
of realistic string interactions in network evolution models with string junctions.
\end{abstract}


\maketitle   

\section{Introduction}
In the context of grand unified theories, phase transitions followed
by spontaneously broken symmetries may leave behind cosmic
strings~\cite{Kibble, Vilenkin_shellard, Hind_Kibble, ms-cs07}, as false
vacuum remnants.  Cosmic strings are generically formed at the end of
hybrid inflation~\cite{shafvil1,shafvil2,jrs03,ms-cs08}. In addition, brane interactions
in the context of string theoretic cosmological models, can
lead~\cite{BMNQR,ma,sarangi-tye,jonesetal,dvalietal} to fundamental
(F) strings, one-dimensional Dirichlet branes (D-strings) and their
bound (FD) states, collectively known as cosmic
superstrings~\cite{Witten,PolchRevis,Sakellariadou:2008ie}, which may
play a cosmological r\^ole as cosmic strings. In particular, cosmic
superstrings are copiously formed at the end of brane
inflation~\cite{dvali-tye99,BMNQR,KKLMMT,delicate}.

Unlike ordinary abelian field theory strings which can only interact
through intercommutation and exchange of partners with probability of
order unity~\cite{ShellProb}, collisions of cosmic superstrings
typically happen with smaller probabiliites and can lead to the
formation of Y-junctions\footnote{This property is also shared by
  non-abelian field theory strings.} at which three strings
meet~\cite{PolchProb,cmp,JJP}. This characteristic property is of
particular interest because it can modify dramatically the network
evolution~\cite{Sakellariadou:2004wq,TWW,NAVOS,Davis:2008kg,PACPS} leading to
potentially observable phenomenological signatures, and thus providing
a potential window into string
theory~\cite{PolchRevis,Davis:2008kg,CPRev,HindmRev,ACMPPS}.

The effect of junctions on the evolution of string networks was the
central subject of several
numerical~\cite{Rajantie:2007hp,SACKdPS,Urrestilla:2007yw,Sakellariadou:2008ay,Bevis:2009az}
and
analytical~\cite{Sakellariadou:2004wq,TWW,NAVOS,Copeland:2006eh,Davis:2008kg,Copeland:2006if,Copeland:2007nv,ACconstr,PACPS}
investigations. In particular, in
Refs.~\cite{Copeland:2006eh,Copeland:2006if,Copeland:2007nv}
and~\cite{Bevis:2008hg} the authors studied in detail the kinematics
of junction formation, under the assumption\footnote{Remarkable
  agreement between the Nambu-Goto and field theory descriptions in
  this context has been demonstrated in
  Refs.~\cite{SACKdPS,Bevis:2008hg,Bevis:2009az}.}  that string
dynamics is well-described by the Nambu-Goto action, and were able to
find kinematic conditions under which junctions form.  These kinematic
constraints were later incorporated into network evolution modelling
in Refs.~\cite{ACconstr,PACPS} and were shown to play an important
r\^ole in determining the relative number densities of the dominant
string species, thus affecting quantitatively any potential
observational signals from these networks~\cite{PACPS,ACMPPS}.
However, an additional potentially significant source of uncertainty
in this type of network models remains, as it is not well-understood
under which conditions these junctions continue to grow and stabilise,
or alternatively shrink resulting in ``unzipping" of the heavier
(bound) string states.  Indeed, the simulations of
Refs.~\cite{Rajantie:2007hp,Urrestilla:2007yw}, studying field theory
models in which zipping can occur, have found evidence supporting that
heavier bound states can actually unzip, leading to a lower abundance
of heavy strings in the network than what one would naively expect.

The purpose of this paper is to study the dynamics of junctions in a
Nambu-Goto approximation and investigate the conditions which could lead
to spontaneous unzipping of the heavier, composite string states.  We
do this by assigning to the junction a mass, thus allowing for cases
in which the tensions of the three strings joining at the junction are
not balanced, and we study the dynamics of the junction under the
influence of these tensions. We find that the straight string configurations under consideration, i.e. with a massive junction that allows for a non-trivial force, are stable under perturbations. More specifically, they exhibit a damped oscillating behaviour around the balance condition solution. Therefore unzipping does not occur and we need to allow for extra forces exerted on the junction. Considering such forces originating from monopole and string forces we indeed find decelerating solutions in the case of straight strings. In the case of loops, unzipping generically occurs as a result of local curvature near the junction.

This paper is organised as follows: In Section~\ref{review}, we
present a brief overview and update of the currently known Y-junction
configurations and their kinematics within the Nambu-Goto
approximation; we pay special attention to the local geometry of
strings near the junction. In Section~\ref{sec:NGeom}, we outline the
basic formalism describing the dynamics of Y-type configurations of
three Nambu-Goto strings ending at a massive junction. We introduce the general setup and obtain the
evolution equations for the junctions. 
In Section~\ref{stability},
we study the kinematics of junction formation for two moving straight
string segments colliding at angle $\alpha$, following
Refs.~\cite{Copeland:2006eh,Copeland:2006if,Copeland:2007nv,Bevis:2008hg}.
We pay particular attention to the local angle, say $\beta$, at the (massive) junction 
and move on to study its dynamics in the formalism of Section~\ref{sec:NGeom}.  
We find that the evolution of $\beta$ is given by damped oscillations
around the equilibrium (critical) value $\beta_{\rm crit}$ (found in
Refs.~\cite{Copeland:2006eh,Copeland:2006if,Copeland:2007nv}) for
which the vector sum of the three \emph{effective} tensions vanishes.  Thus, even if we deform the tension balance condition, the local
angle quickly relaxes to $\beta_{\rm crit}$ and the junction grows
relativistically, as has been assumed in most string evolution models.
In Section~\ref{unzipping}, we investigate other possible unzipping mechanisms, namely monopole forces, string forces, and string curvature for loops with junctions. We find solutions which exhibit decelerated zipping, eventually leading to the unzipping of the Y-junction configuration. We round up our
conclusions in Section~\ref{sec:concls}.

\section{Massless String Junctions: Review and Update\label{review}}

\begin{figure}[h!]
\centering \includegraphics[scale=0.7]{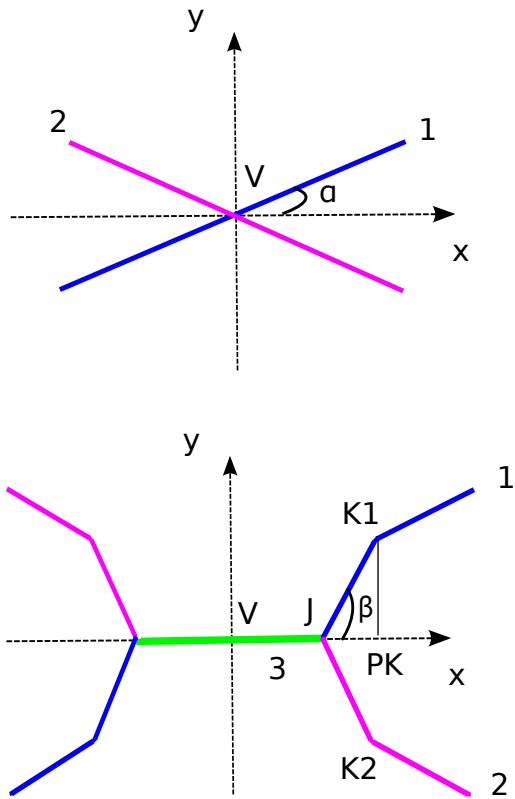}
\caption{\textit{Top}: Two strings (denoted by 1, 2) colliding at time
  $t=0$. \textit{Bottom}: At time $t=\delta t$, a new segment (denoted
  by 3) has formed. The junction $J$ is moving to the right and the
  zipper is growing. }
\label{unzipconfig}
\end{figure}

Let us start by reviewing the kinematic condition for junction
formation derived in Ref.~\cite{Copeland:2006eh}, hereafter
CKS. The authors considered a configuration of two straight infinite
string segments with tensions $\mu_1$ and $\mu_2$, moving towards one
another each with speed $v$ and colliding at an angle $2\alpha$ at
time $t=0$.  Choosing coordinates such that the two strings collide in
the $(x,y)$-plane, each at an angle $\pm \alpha$ off the $x$-axis
(Fig.\ref{unzipconfig}, {\it Top}), the two strings in this
configuration for $t\le 0$ can be parametrised in terms of two
parameters $(\sigma,t)$ as 
\be\label{stringsatalpha} \bx_{1,2} = (-\sigma \gamma_v^{-1} \cos
\alpha, \mp \sigma \gamma_v^{-1} \sin \alpha, \pm v t)~, \ee
with $\gamma_v^{-1}=\sqrt{1-v^2}$. This is a solution of the
Nambu-Goto equations of motion expressed in the conformal temporal gauge, 
where the spacelike worldsheet parameter $\sigma$ is the {\it invariant length}, 
${\rm d}\sigma={\rm d}|\bx|/\sqrt{1-\dot\bx^2}$, and
the timelike worldsheet parameter is identified with background time
$t$. Note that $\dot\bx_{1,2}=(0,0,\pm v)$ is
the physical velocity, which is transverse to the string tangent
$\bx^{\prime}\equiv \partial_\sigma \bx$, and since $v$ is constant
the quantity $|\sigma| \gamma_v^{-1}$ in Eq.~(\ref{stringsatalpha}) simply
measures the {\it physical length} along the string. In the chosen
sign convention $\sigma$ increases towards the vertex.

As a result of the collision at $t=0$, a new string segment (a ``link"
or ``zipper") of tension $\mu_3$ can be formed, giving rise to two
trilinear Y-shaped junctions connecting it to the original strings
(Fig.\ref{unzipconfig}, {\it Bottom}).  Far away from the junctions
the strings retain their original motion and orientation, so this
2-junction configuration is connected to the solution
(\ref{stringsatalpha}) for $t>0$ through four kinks, moving at the
speed of light along the original strings and away from the point of
string intersection. Here, we concentrate on the simplest case, where
the original strings have equal tension $\mu_1=\mu_2\equiv
\mu$. Symmetry then implies that the newly formed ``zipper" segment
lies either along the $x$-axis or along the $y$-axis and that it stays
in the $(x,y)$-plane at all times $t>0$. For small angle $\alpha$ we
may expect the zipper to be formed along the $x$-axis, as shown in
Fig.\ref{unzipconfig}, {\it Bottom}.

The authors of Ref.~\cite{Copeland:2006eh} were able to confirm
and quantify the last statement. By writing down and analysing the
Nambu-Goto action for three strings $\bx_i(\sigma_i,t)$; $i\in
\{1,2,3\}$, joining at a vertex $X(t)$, they were able to find the
kinematic conditions that must be satisfied for junction formation to
be possible in the configuration (\ref{stringsatalpha}). A first
restriction is that the string tensions must satisfy the triangle
inequalities. Further, it is possible to obtain a quantitative
constraint depending explicitly on the tensions, the speed $v$ and
angle $\alpha$. This can be done by thinking of the point of
intersection at $t=0$ (Fig.\ref{unzipconfig}, {\it Top}) as a
``zipper" of tension $\mu_3$ but zero length, which can grow for $t>0$
resulting in the configuration of Fig.~\ref{unzipconfig}, {\it Bottom}.
Let $s_i(t)$ be the coordinate of the junction on string $i$, that is
$\bx_i(s_i(t),t)=X(t)$ for all three strings.  Then, for
$\mu_1=\mu_2=\mu$, CKS found
\be\label{constr}
\dot s_3 = \frac{2\mu \gamma_v^{-1} \cos\alpha - \mu_3}{2\mu
-\mu_3\gamma_v^{-1}\cos\alpha} \,.
\ee    
The constraint for junction formation is that the zipper must grow, 
$\dot s_3>0$, which gives
\be\label{xconstr}
\alpha < \arccos\left(\frac{\mu_3\gamma_v}{2\mu}\right)\,. 
\ee    
This defines the region of parameter space for which formation of 
a zipper along the $x$-axis is kinematically allowed. A similar constraint 
can be obtained for a zipper along the $y$-axis, which is expected to be 
the preferred configuration for large angles $\alpha<\pi/2$. Indeed, in 
this case one finds
\be\label{yconstr}
\alpha > \arcsin\left(\frac{\mu_3\gamma_v}{2\mu}\right)\, , 
\ee 
confirming the above expectation.  

The result (\ref{xconstr}-\ref{yconstr}) has important implications for the evolution 
of string networks with junctions, where one must consider collisions at all possible 
angles and velocities in a large ensemble of string segments. It tells us that string 
junctions only form in a subset of orientations $\alpha\in [0,\pi/2]$ defined by the union 
of the regions (\ref{xconstr})-(\ref{yconstr}). In addition, there is a critical speed, 
depending on the ratio of the tensions, beyond which the junction cannot be formed. Indeed, 
(\ref{xconstr})-(\ref{yconstr}) are well-defined for $\gamma_v\le 2\mu/\mu_3$, whose saturation 
corresponds to a maximum speed. Although these results have been derived in the 
zero-width, Nambu-Goto approximation, they were found to be in remarkable agreement with 
Abelian-Higgs field theory simulations of straight string collisions~\cite{SACKdPS}.  Thus, 
these constraints, as well as their generalisation for unequal tensions~\cite{Copeland:2006if} 
and cosmic superstrings~ \cite{Copeland:2007nv}, must be (and have been) incorporated in 
network evolution models, significantly affecting quantitative predictions~\cite{ACconstr,PACPS}.           

Equation (\ref{constr}) was derived by considering the string
configuration (\ref{stringsatalpha}) at $t=0$, when the angle between
the colliding segments and the $x$-axis at the junction is $\pm
\alpha$. One may ask whether this equation is valid for $t>0$ when,
clearly, the angle between the strings and the $x$-axis at the
junction (segments $JK_1$, $JK_2$ in Fig.~\ref{unzipconfig}, {\it
  Bottom}) is $\pm \beta$, with $\beta>\alpha$. In fact,
Eq.~(\ref{constr}) is actually valid for all $t>0$, as the solution
has $\dot s_1=\dot s_2=-(\mu_3/2\mu)\dot s_3={\rm const}$. For
$t>0$, Eq.~(\ref{constr}) can be understood by considering the union
of segments $JK_1$ and $JK_2$ as a ``rigid" body subject to the
tension $\mu_3$ of the ``zipper" segment $VJ$ (applied on point $J$ and
pulling to the left) and the tensions of the two strings beyond the
kinks (which are applied at points $K_1$, and $K_2$ at angles $\pm
\alpha$ with the $x$-axis respectively) together pulling to the
right. This was done in detail in Ref.~\cite{Bevis:2008hg}. The
authors of Ref.~\cite{Bevis:2008hg} showed that the growth of the
segments $JK_1$ and $JK_2$ leads to a rate of change of $x$-momentum:
\be\label{pxdot} 
\dot p_x = \dot s_3\left(2\mu-\mu_3 \gamma_v^{-1}
\cos\alpha \right) \,, 
\ee which is exactly balanced by the
$x$-component of the sum of the external tensions:
\be\label{xsumtensions} T_x = 2\mu \gamma_v^{-1} \cos\alpha - \mu_3\,.  
\ee 
Equating Eq.~(\ref{pxdot}) and
Eq.~(\ref{xsumtensions}) we recover Eq.~(\ref{constr}), which is a
statement of energy-momentum conservation.

Thus, in the solution considered, where $\dot s_3={\rm const}$, the
``local" angle $\beta={\rm const}>\alpha$ and can be eliminated from
the dynamics. The resulting configuration is an ever-growing zipper,
$\dot s_3>0$.  Motivated by numerical
simulations~\cite{Rajantie:2007hp,Urrestilla:2007yw} which suggest
that the zipper growth can be inverted in string networks (thus leading to
string unzipping), we are interested in studying deviations from this
model solution, allowing, in particular, non-trivial evolution of the
local angle $\beta$. This will be done in the following sections, by
considering perturbations around this solution and by introducing a
massive junction, subject to external forces, allowing us to study
junction dynamics. In the remaining of this section we will complete
the above basic picture by describing the configuration of
Fig.~\ref{unzipconfig} near the junction, in terms of the angle
$\beta$. Starting with the simplest possible geometric configuration,
we will first assume that $v\rightarrow 0$ as $t\rightarrow 0$,
ensuring that all strings stay on the $(x,y)$-plane for all $t\ge
0$. Once this configuration is fully described, we will then restore
the $z$-velocity, $v$.

By symmetry it suffices to consider only one-half of
Fig.~\ref{unzipconfig}, e.g. $x>0$, the other half being just the
mirror image. First, we note that the angle $\beta$ can be easily
determined from $\dot s_3$ and the fact that the kinks move at the
speed of light at angles $\pm \alpha$ off the $x$-axis. In fact, even
if we did not know that $\dot s_3={\rm const}$, we could still
determine $\beta$ for sufficiently small time $\delta t>0$ such that
$s_3(\delta t)\simeq \dot s_3(0) \delta t$. This only assumes continuity
of $\dot s_3(t)$ at $t=0$, where $\dot s_3(0)$ is determined from
Eq.~(\ref{constr}).  We are also assuming that the segments $JK_1$, $JK_2$
are straight.

Let us determine $\beta$ in the configuration shown in the bottom panel 
of Fig.~\ref{unzipconfig}.  Since $\dot s_3(0)>0$, the junction $J$ is
moving to the right and the zipper is growing.  As mentioned above,
causality requires that sufficiently far from the junction the
solution (\ref{stringsatalpha}) is still valid at time $t=\delta t$
and this solution is joined to the segments $JK_1$, $JK_2$ at the
moving kinks, $K_1$, $K_2$.  Due to longitudinal Lorentz invariance
the kinks move along the strings at speed $c=1$ and so they are
positioned at a distance $c\delta t=\delta t$ away from the original vertex
$V$ and along the original direction of the strings, i.e. at angles
$\pm \alpha$.  The junction $J$ only moves at speed
$\dot s_3<1$ along the $x$-axis so it is at a distance $\dot
s_3(0)\delta t$ from the original vertex. This is the half-length of
the zipper at $t=\delta t$ (the other half is in the mirror half of
Fig.~\ref{unzipconfig} {\it Bottom}, i.e., the one with $x<0$, and
involves a junction moving to the left):  
\[
\frac{\ell_3(t=\delta t)}{2}=VJ=\dot s_3(0)\delta t \,.
\]             
Overall, the configuration at $t=\delta t$ involves the zipper segment $VJ$ 
(with tension $\mu_3$), two straight segments (with equal tensions $\mu$) 
linking the junction $J$ to the kinks $K_1$ and $K_2$, plus
the original string segments labeled by 1 and 2 beyond $K_1$ and
$K_2$, respectively.  The angle $\beta$ between $JK_1$
and the $x$-axis is given by (see, Fig.~\ref{unzipconfig}, {\it Bottom})
\be\label{cosbeta}
\cos \beta=\frac{VK_1\cos \alpha -VJ}{JK_1}=\frac{[\cos\alpha-\dot
s_3(0)]\delta t}{\sqrt{\delta t^2+\dot s_3(0)^2\delta t^2 -2\dot
s_3(0)\cos{\alpha}\delta t^2}}= \frac{\cos\alpha-\dot
s_3(0)}{\sqrt{1+\dot s_3(0)[\dot s_3(0)-2\cos{\alpha}]}}\,.
\ee
Similarly, we also have (directly from Fig.~\ref{unzipconfig}, {\it Bottom})
\be\label{sinbeta}
\sin\beta=\frac{K_1 P_K}{JK_1}=
\frac{\sin\alpha}{\sqrt{1+\dot s_3(0)[\dot s_3(0)-2\cos{\alpha}]}}\,,
\ee  
and
\be\label{tanbeta}
\tan\beta=\frac{K_1 P_K}{VK_1\cos\alpha-VJ}=
\frac{\sin\alpha}{\cos\alpha-\dot s_3(0)}\,.
\ee
Since $\dot s_3(0)$ is positive, and also\footnote{Physically, this
  guarantees that the junction does not move to the right faster than
  the projection $P_K$ of the kinks $K_1$, $K_2$ on the x-axis.} $\dot
s_3(0)\le \cos\alpha$ from (\ref{constr}) above, we have
\[ 
\tan\beta > \tan\alpha \,.
\]    
Thus, for any $\alpha<\arccos(\mu_3/2\mu)$, Eq.~({\ref{tanbeta})
  implies that the angle at the junction must change discontinuously
  from $\alpha$ to $\beta>\alpha$ at $t=0$, when $\dot s_3>0$ is
  suddenly switched on. For $\alpha=\arccos(\mu_3/2\mu)$ we have $\dot
  s_3(0)=0$ so the zipper does not form. It stays formally at zero
  length and the junction does not move, which is expected since
  $\alpha=\arccos(\mu_3/2\mu)$ corresponds to the balance of tensions
  at the junction, Eq. (\ref{xsumtensions}). In the other extreme, 
  $\alpha=0$, we get $\dot s_3(0)=1$ and the junction moves to the right at the speed of light.

Using Eq.~(\ref{constr}) we can re-write Eq.~(\ref{tanbeta}) as 
(recall we are assuming $v\rightarrow 0$ as $t\rightarrow 0$ from below)
\be\label{tanbeta1}
\tan\beta=
\frac{(2\mu/\mu_3)-\cos\alpha}{\sin\alpha} \,. 
\ee
For $\alpha\rightarrow 0$ we get $\beta\rightarrow \pi/2$, while for
$\alpha=\arccos(\mu_3/2\mu)$ we find also that
$\beta=\arccos(\mu_3/2\mu)$. In other words,
$\beta\in[\arccos(\mu_3/2\mu),\pi/2]$, as we may have expected from
the geometry.  This may at first appear counterintuitive ---
especially the statement that $\alpha=0$ produces $\beta=\pi/2$ ---
but the picture is clear: $\alpha=0$ corresponds to the two strings
being aligned, which as we saw gives velocity $\dot s_3=1$ for the
junction.  However, the kink projection velocity $\cos\alpha$ for
$\alpha=0$ is also 1 (the kink is trivial and moves along the
$x$-axis).  Thus, both the kink and the junction move at the same
speed along the $x$-axis, starting at the same point $V$. Formally,
the segments $JK_1$ and $JK_2$ are at right angle to the $x$-axis, but
they have zero length.  More generally, small $\alpha$ leads to
$\beta$ near $\pi/2$.  In the other extreme,
$\alpha=\arccos(\mu_3/2\mu)$, we have $\dot s_3=0$ as explained above
and the zipper does not grow, staying formally at zero length.  The
``kinks'' are trivial and propagate up the original strings: there is
no bend so $\beta=\alpha=\arccos(\mu_3/2\mu)$.
    
A source of potential confusion is that Eq.~(\ref{constr}) is clearly
not satisfied for the angle $\beta$, yet the configuration in
Fig.\ref{unzipconfig}, {\it Bottom}, appears to be identical to the
one used for the derivation of Eq.~(\ref{constr}).  As we saw, to derive
Eq.~(\ref{constr}) one takes the $x>0$ half of the configuration of
Fig.~\ref{unzipconfig}, {\it Top}, and considers the vertex $V$ as a
zipper of tension $\mu_3$ and zero length at $t=0$. The difference
from the configuration in Fig.~\ref{unzipconfig}, {\it Bottom}, appears
to be the length of the zipper (zero vs finite length) and the angle
at the junction ($\alpha$ vs $\beta$). How can it be that the same
equation (\ref{constr}) is not valid for the angle $\beta$? The answer
is that in the configuration of Fig.~\ref{unzipconfig}, {\it Top}, the
string segments only have velocity $v$ in the $z$-direction (even
though we are taking $v\rightarrow 0$ in this simplest example), while
in that of Fig.~\ref{unzipconfig}, {\it Bottom}, the junction is moving
with velocity $\dot s_3$ to the right so the segments $JK_1$, $JK_2$
must have non-zero transverse velocity, $w$, in the
$(x,y)$-plane. Thus, the solution (\ref{stringsatalpha}) used to
derive Eq.~(\ref{constr}) for the angle $\alpha$ does not apply to the
configuration in Fig.~\ref{unzipconfig}, {\it Bottom}. The correct
parametrisation of this configuration (up to the kinks $K_1$, $K_2$)
for $t>0$ is
\be\label{stringsatbeta}
\begin{array}{rclcl}
\bx_1 &=& (-\sigma \gamma_w^{-1} \cos\beta + w \sin\beta\, t, - \sigma
\gamma_w^{-1} \sin\beta - w\cos\beta\, t, 0) &,& \; -t \le\sigma\le
s_1(t) \\ \bx_2 &=& (-\sigma \gamma_w^{-1} \cos\beta + w
\sin\beta\, t, + \sigma \gamma_w^{-1} \sin\beta + w\cos\beta\, t, 0)
&,& \; -t \le\sigma\le s_2(t) \\ 
\bx_3 &=& (\sigma, 0, 0) &,& \;\;\; 0 \le \sigma \le s_3(t) 
\end{array}
\ee        
which is a solution of the Nambu-Goto equations in the conformal
gauge. Note that for strings 1 and 2 (segments $JK_1$ and $JK_2$) we
have only kept the physical {\it transverse} velocities, which
corresponds to our choice of conformal gauge.  The magnitude $w$ of
the transverse velocities for the two strings can be readily found to
be $w=\dot s_3 \sin\beta$. Using this solution for $\bx_1$, $\bx_2$
and repeating the analysis of Ref.~\cite{Copeland:2006eh} we arrive at
the following equation expressing $\dot s_3$ in terms of the local
angle at the junction, $\beta$:
\be\label{dots3beta} \dot s_3 = \frac{ 2\mu ( \gamma_w^{-1} \cos\beta
+ w \sin\beta ) - \mu_3}{2\mu - \mu_3 ( \gamma_w^{-1} \cos\beta + w
\sin\beta )}\,.  \ee
Clearly, $\dot s_3={\rm const}$, consistent with Eq.~(\ref{constr}).
As already mentioned, even though Eq.~(\ref{constr}) was derived from
the solution (\ref{stringsatalpha}), valid only up to $t=0$ but not
later, it holds for all $t>0$. This now becomes clear via the equation
$\cos\alpha=\gamma_w^{-1} \cos\beta + w \sin\beta$, which can be
verified with $w=\dot s_3 \sin\beta$. For $\alpha=\arccos(\mu_3/2\mu)$
we have $w=0$ so $\beta=\alpha$, while in the other extreme,
$\alpha=0$, we have $w=1$, $\beta=\pi/2$.

This complements the analysis of Refs.~\cite{Copeland:2006eh} and
\cite{Bevis:2008hg} for $\mu_1=\mu_2$. Restoring the velocity $v$, the
simple geometrical picture above in which all strings lie on the
$(x,y)$-plane is lost, but it is still straightforward to construct
the solution algebraically. The solution along each of the segments
$JK_1$ and $JK_2$ can be represented in terms of a unit direction
vector and a transverse velocity~\cite{Bevis:2008hg}
\be\label{wdparam} {\bf x}_i = \frac{\sigma}{\gamma_w} {\bf d}_i + t\,
         {\bf w}_i \;\;\; ,\; i\in\{1,2\} \ee
such that ${\bf d}_i\cdot {\bf w}_i=0$. This parametrisation again
corresponds to choosing the conformal gauge, so that $\sigma$ measures
{\it invariant} length. Since the segments $JK_i$ must be joined to
the zipper $VJ$ at the junction $J$ and to the original solution
(\ref{stringsatalpha}) at the kinks $K_1$, $K_2$, the vectors ${\bf
  d}_i$ and ${\bf w}_i$ can be determined by requiring continuity at
$\sigma=s_3(t)$ and $\sigma=-t$, as well as imposing the conditions
${\bf x}_i(-t,t)={\bf x}_3(s_3(t),t)+\vec{JK_i}$.  One then
finds\footnote{We will examine these solutions and their
  generalisation for $\mu_1\ne \mu_2$ in Section~\ref{sec:unzipgeom},
  where we will also allow for massive junctions and study their
  dynamics.} that the segments $JK_1$, $JK_2$ and $VJ$ still lie on a
plane, albeit one which is rotated by an angle
$\tan\phi=v\gamma_v\csc{\alpha}$ around the $x$-axis
(Fig. \ref{Junction3D}). The local configuration at the junction is
still described by the angle $\beta$ discussed above, which for $v\ne
0$ is given by \be\label{cosbetavnonzero} \cos
\beta=\frac{\cos\alpha/\gamma_v-\dot s_3}{\sqrt{1+\dot s_3[\dot
      s_3-2\cos{\alpha}/\gamma_v]}}\,.  \ee

\begin{figure}[h!]
\includegraphics[scale=0.6]{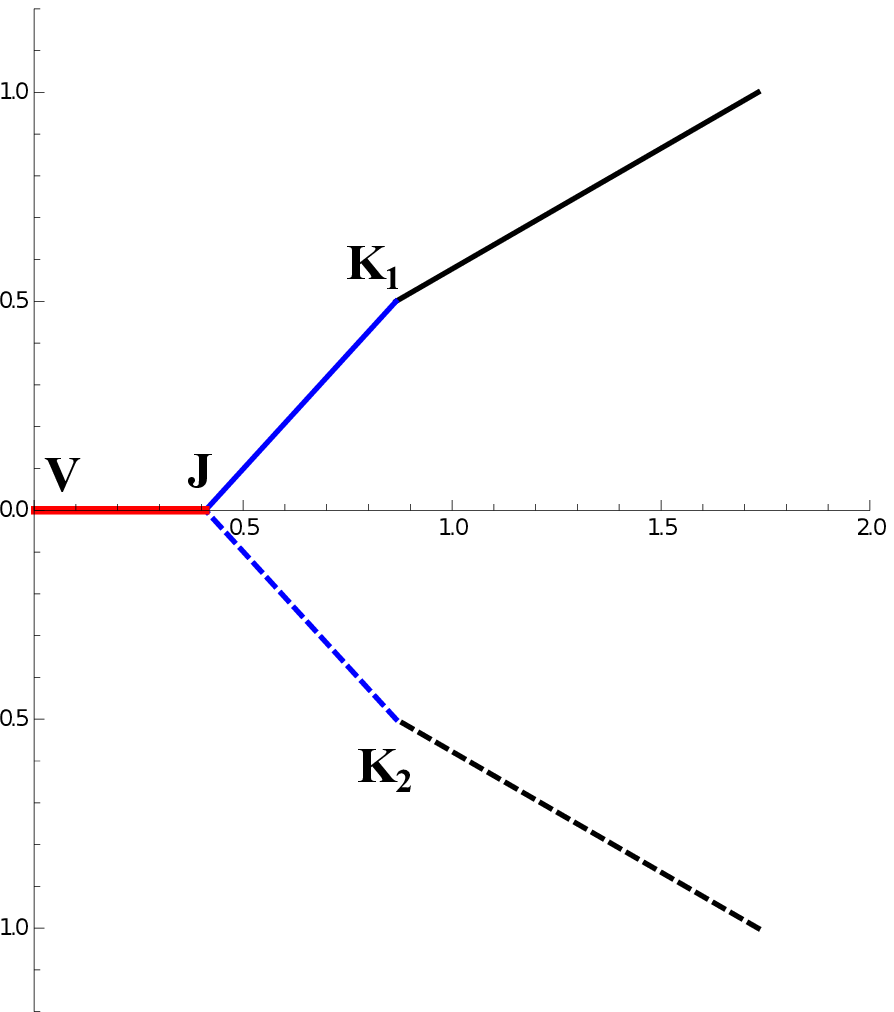}
\includegraphics[scale=1.1]{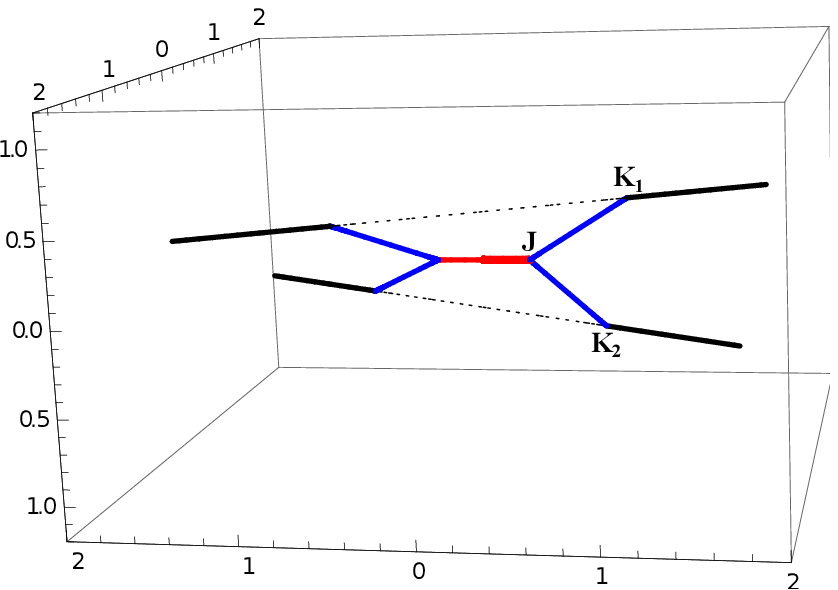}
\caption{String junction configurations for $v\rightarrow 0$ as
$t\rightarrow 0$ ({\it Left}) and for $v={\rm const}>0$ ({\it
Right}).}
\label{Junction3D}
\end{figure}
Focusing on the local structure at the junction in terms of the
geometric picture described, it is clear that for these solutions to
make sense with $\beta=\rm const$, $\dot s_3=\rm const$, the vector
sum of the tensions ${\bf T}_i=\mu_i {\bf x}_i^\prime$ of all strings
at the junction must exactly balance the transfer of momentum due to
the shrinking of the segments $JK_1$ and $JK_2$, which happens for
$\dot s_3>0$.  This is simply a statement of energy conservation. The
momentum carried by a segment of invariant length $L$ and constant
transverse velocity ${\bf u}$ is ${\bf p}=\mu L {\bf u}$, so if the
segment is shrinking then the rate of change of momentum is $\dot {\bf
p}=\mu \dot L {\bf u}$.  At the junction, $s_i(t)$ labels invariant
length so for each string we have $\dot {\bf p}_i=\mu_i \dot
s_i(t)\dot {\bf x}_i$.  Concentrating on the $x$-component we have
\be\label{junctionbalance} \sum (T_x+ \dot p_x) = 2\mu \cos\beta
\gamma_w^{-1} - 2\mu \dot s_1 w_x - \mu_3 = 0 \,, \ee
which can be simplified to
\be\label{junctionbalance1} 2\mu \cos\beta \gamma_w -
\mu_3 = 0 \,.  \ee
In view of the CKS constraints described in the beginning of this
section, it now becomes obvious that the angle $\beta$ is indeed the
{\it critical angle} saturating the analogue of
Eqs.~(\ref{xconstr})-(\ref{yconstr}) for the configuration of
Fig.~\ref{Junction3D}, {\it Left}. This is as expected: $\beta$ is the
unique angle balancing tensions and momentum transfer, which resolves
the original mismatch of tensions at $t=0$.  This configuration of
local equilibrium is the reason why the growth of the zipper $\dot
s_3$ can be described by considering the external tensions acting on
the union of $JK_1$ and $JK_2$ thought of as a ``rigid" body,
Eqs.~(\ref{pxdot})-(\ref{xsumtensions}).

Our motivations for studying the local dynamics of string junctions
can be rephrased in this context in terms of studying the stability of
the solutions just described. In particular, we are interested in
understanding the conditions under which non-trivial dynamics of
string junctions could allow for the local equilibrium conditions to
evolve so as to decelerate (and potentially invert) the zipping
process. More generally, we wish to explore possible mechanisms for the
unzipping of string junction configurations, arguably seen in
some field theory simulations. In the next section we move on to the
general formalism for studying the dynamics of massive junctions.

\section{Massive Junctions: Nambu-Goto action and dynamics}
\label{sec:NGeom}

In this section, we generalise the action of
Ref.~\cite{Copeland:2006if} by including a \emph{massive} junction and
derive the equations of motion.  We are working with the configuration
shown in Figs.~\ref{unzipconfig} and \ref{Junction3D}, concentrating
on the $x>0$ half $x$-axis. We parameterise the position of the $i$-th
string ($i=1,2,3$) as
\be x^\mu_i(\tau, \sigma_i)~, \ee 
where $\tau$ and $\sigma_i$ are the world-sheet coordinates; the timelike 
coordinate $\tau$ is chosen to be the same for all three strings. The induced
metric on the world-sheet for the string $i$ is 
\be
\gamma^i_{ab}=\frac{\partial x^\mu_i}{\partial \sigma^a}\frac{\partial
  x^\nu_i}{\partial \sigma^b}\eta_{\mu\nu}~, \ee
where $\sigma^a=(\tau,\sigma_i)$ and $\eta_{\mu\nu}$ is the Minkowski metric
with signature $(+,-,-,-)$.  A dot/dash denotes differentiation 
with respect to $\tau/\sigma_i$, respectively. The values of the
world-sheet coordinate $\sigma_i$ at the junction are denoted by $s_i$
and are generally $\tau$-dependent. We choose $\sigma_i$ to
increase towards junction $J$ for all three strings. The position of
the junction with mass $m$ is 
\be \nonumber
X^\mu_m(\tau)=x^\mu_i(\tau,s_i(\tau))~~~,~~~ \mbox{with}~~~~ i=1,2,3~. \ee 
We are working in the conformal gauge, where 
\be
\label{confgauge}
\gamma^i_{\tau\tau}+\gamma^i_{\sigma_i\sigma_i}=0 \, \, \, \, ; \, \, \, \, 
\gamma^i_{\tau\sigma_i}=0~.
\ee
The Nambu-Goto action for the three strings of tensions $\mu_i$
(with $i=1,2,3$) meeting at junction $J$ with mass $m$ is
\bea \nonumber
\label{action}
S&=&-\displaystyle\sum_i      \mu_i      \int     d\tau      d\sigma_i
\Theta(s_i(\tau)-\sigma_i)     \sqrt{-x'^2_i\dot{x}^2_i}     \\    &&+
\displaystyle\sum_i       \int        d\tau       f_{i\mu}       \cdot
                 [x^\mu_i(\tau,s_i(\tau))-X^\mu_m(\tau)]-m\int    d\tau
                 \sqrt{\dot{X}^2_m}~.  \eea
Varying the above action, Eq.~(\ref{action}), with respect to $x^{\mu}_i$
yields the usual equation of motion for a string in Minkowski
space-time (away from the junction), which is the wave equation 
\be
\label{wave-eq_1}
\ddot{x}^\mu_i-x^{\mu\prime\prime}_{i}=0\,.  
\ee
The boundary terms, i.e. the ones proportional to
$\delta(s_i(t)-\sigma_i)$, give
\be
\label{boundaries}
\mu_i(x^{\mu\prime}_i+\dot{s}_i\dot{x}^\mu_i)=-f^\mu_i\,,
\ee
where the functions are evaluated at $(\tau,s_i(\tau))$. Varying the
Lagrange multipliers $f^\mu_i$ provides the boundary condition
\be
\label{boundary-condition}
X^\mu_m(\tau)=x^\mu_i(\tau,s_i(\tau))~,  \ee
and varying $X^\mu_m$ gives 
\be
\label{varXm}
m\frac{d}{d\tau}\left(\frac{\dot{X}^\mu_m}{\sqrt{\dot{X}^2_m}}\right)
=\displaystyle\sum_if^\mu_i~.  \ee
Using Eqs.~(\ref{boundaries}) and (\ref{varXm}), we obtain 
\be
\label{eomXm}
m\frac{d}{d\tau}\left(\frac{\dot{X}^\mu_m}{\sqrt{\dot{X}^2_m}}\right)
=-\displaystyle\sum_i\mu_i(x^{\mu\prime}_i+\dot{s}_i\dot{x}^\mu_i)~.
\ee
Let us impose the temporal gauge condition $x^0\equiv t=\tau$. The conformal
gauge conditions, Eq.~(\ref{confgauge}), then reduce to 
\be
\label{gauge}
\dot{\bx}^2_i+\bx^{\prime 2}_i=1\, \, \, \, ;\, \, \, \, \dot{\bx}_i
\cdot \bx^{\prime}_i=0~, 
\ee
where $\bx_i$ is the spatial part of $x^{\mu}_i$, so that $x^{\mu}_i=(t,\bx_i)$. 
The 4-dimensional wave equation, Eq.~(\ref{wave-eq_1}), reduces to the 3-dimensional 
wave equation
\be
\ddot{\bx}_i-\bx^{\prime\prime}_i=0~, \ee
with solution 
\be
\bx_i(\sigma,t)=\frac{1}{2}[\mathbf{a}_i(\sigma+t)+\mathbf{b}_i(\sigma-t)]~,
\ee
and the gauge conditions in Eq.~({\ref{gauge}) imply\footnote{Note
    that, when applied to $\mathbf{a}$ and $\mathbf{b}$, the prime
    denotes derivatives with respect to their arguments $(\sigma+t)$
    and $(\sigma-t)$, respectively.}
\be \mathbf{a}^{\prime2}_i=\mathbf{b}^{\prime2}_i=1\,.
\ee 
The equation obtained from the boundary terms,
Eq.~({\ref{boundaries}), becomes
\be
\mu_i(\bx^{\prime}_i+\dot{s}_i\dot{\bx}_i)=-\mathbf{f}_i~, \ee
where the functions are evaluated at $(t,s_i(t))$.  Moreover, the
boundary condition, Eq.~(\ref{boundary-condition}), simplifies to
\be
\bx_i(t,s_i(t))=\bX_m(t)~, \ee
which gives 
\be
\label{dotbXm}
\dot{\bX}_m=\bx^\prime_i\dot{s}_i+\dot{\bx}_i~.
\ee
Let us consider Eq.~(\ref{eomXm}): its 0-th component implies the
energy conservation equation 
\be
\label{energycons}
m\dot{\gamma}_m+\displaystyle\sum_i \mu_i
\dot{s}_i=0~~~~\mbox{where}~~~~\gamma_m=\frac{1}{\sqrt{1-\dot{\bX}^2_m}}~,
\ee
and its $i$-th components lead to
\be
\label{ddotXm1}
m\frac{d}{dt}\left(\gamma_m \dot{\bX}_m\right)=-\displaystyle\sum_i
\mu_i(\bx^{\prime}_i+\dot{s}_i\dot{\bx}_i)~.  \ee
The above equation, Eq.~(\ref{ddotXm1}), can be written as 
\be
\label{eq19}
m\dot{\gamma}_m\dot{\bX}_m+m\gamma_m\ddot{\bX}_m=
-\displaystyle\sum_i\mu_i(\bx^{\prime}_i+\dot{s}_i\dot{\bx}_i)~,  \ee
and using Eqs.~(\ref{dotbXm}) and (\ref{energycons}) we get
\be\label{Xddotsi}
m\gamma_m\ddot{\bX}_m=-\displaystyle\sum_i\mu_i(1-\dot{s}^2_i)\bx^\prime_i~.
\ee
Finally, using the gauge conditions, Eqs.~(\ref{gauge}), and Eq.~(\ref{dotbXm}) above, we obtain
\be
\label{ddotXm2}
\ddot{\bX}_m=-\frac{1}{m}\gamma^{-3}_m\displaystyle\sum_i\mu_i\frac{\bx^\prime_i}{\bx^{\prime
    2}_i}~.  \ee
Equations~(\ref{energycons}) and (\ref{ddotXm2}) agree with the
analogous equations found in Refs.~\cite{Martin:1996cp,
  Siemens:2000ty} for a system of monopoles connected to two strings
each.

Note that, using the gauge conditions, Eq.~($\ref{gauge}$), we can write
Eq.~(\ref{dotbXm}) as
\be
\label{dotseqv1}
\dot{s}_i(t)=\frac{\dot{\bX}_m(t) \cdot \bx'_i(s_i(t),t)}{|\bx^{\prime
    2}_i(s_i(t),t)|}~.  \ee
Since
\be
\bX_m=\bx_i(s_i(t),t)=\frac{1}{2}[\bb_i(s_i(t)-t)+\ba_i(s_i(t)+t)]~,
\ee
the vertex velocity can be written as
\be
\dot{\bX}_m=\frac{1}{2}[-(1-\dot{s}_i(t))\bb'_i+(1+\dot{s}_i(t))\ba'_i]~.
\ee
We can therefore express the outgoing waves, $\ba'_i(s_i(t)+t)$, as a
function of the incoming waves, $\bb'_i(s_i(t)-t)$, and the vertex
velocity, $\dot{\bX}_m$, in the following way:
\be
\label{aprimeeq}
\ba'_i(s_i(t)+t)=\frac{2\dot{\bX}_m+(1-\dot{s}_i)
\bb'_i(s_i(t)-t)}{1+\dot{s}_i}~.
\ee
Starting from Eq.~(\ref{dotseqv1}) and using Eq.~(\ref{aprimeeq}) and
$\bx'=(\bb'+\ba')/2$ with $\bb'^2=\ba'^2=1$, we obtain
\be
\label{dotseqv2}
\dot{s}_i(t)=\frac{\dot{\bX}^2_m(t)+\dot{\bX}_m(t) \cdot
  \bb'_i(s_i(t)-t)}{\dot{\bX}_m(t) \cdot \bb'_i(s_i(t)-t)+1}~.  \ee
We have thus obtained the general evolution equations for 3 strings of
different tensions ending on a massive junction, and have derived the
equivalent to Eqs.~(\ref{constr}) and (\ref{dots3beta}). Note that a
general result arising from these equations is a co-planar condition
on ${\bf x}_i^\prime$ and $\ddot{\bf X}$.  Indeed, from
Eq. (\ref{Xddotsi}) we get
\be
\label{co-planar}
m\gamma_{\rm m}\ddot{\bX}_{\rm m} \cdot (\bx'_2 \times \bx'_3)=
-\mu_1 (1-\dot{s}^2_1)\bx'_1 \cdot (\bx'_2 \times \bx'_3)\,,
\ee
which is satisfied with $\ddot{\bX}_{\rm m}$ and $\bx'_i$ co-planar at
the point of the junction\footnote{Note that in the massless junction
  case, one finds $\bx'_1 \cdot (\bx'_2 \times \bx'_3)=0\,,$ which
  indicates that the $\bx'_i$ are co-planar at the point of the
  junction~\cite{Copeland:2007nv}.}.

\section{Stability of Y-junction configurations}
\label{stability}

Clearly, all configurations considered in Section~\ref{review} are
special solutions of the equations of the previous section with $\ddot
{\bf X}_m=0$. But some of the features we saw in those special
solutions are also present in the most general Y-type
configurations. For example, the joining strings are co-planar in the
vicinity of the junction by virtue of Eq.~(\ref{co-planar}). Let us
briefly consider the generalisation of our discussion of the balance
between tension and momentum transfer in
Eqs.~(\ref{junctionbalance})-(\ref{junctionbalance1}). This is
described by Eq.~(\ref{ddotXm1}), where $\mu_i\bx_i^\prime$ and
$\mu_i\dot s_i \dot\bx_i$ are the vector tension and rate of change of
momentum for the $i$-th string. They are now allowed to be unbalanced,
resulting in acceleration $\ddot{\bf X}_m$ of the massive
junction. Using our gauge constraints, this equation is equivalent to
Eq.~(\ref{ddotXm2}) which is remarkably simple: for any straight string
with velocity $w$ we have that $|\bx'|=\gamma_w^{-1}$ so the overall
effect of adding the momentum transfer can be effectively described by
rescaling the tension from $T=\mu\gamma_w^{-1}$ to $T_{\rm
  eff}\equiv\gamma_w^2T=\mu\gamma_w$, which is what we found in
Eq.~(\ref{junctionbalance})-(\ref{junctionbalance1}) for the
$x$-component. The sum of these effective tensions at the junction is
then proportional to $\ddot {\bf X}_m$, according to 
Eq.~(\ref{ddotXm2}).

From this simple picture we may then expect the Y-junction
configuration to be stable under small perturbations of the
angle. Focusing on our familiar example $\mu_1=\mu_2\equiv\mu$,
Eq. (\ref{ddotXm2}) becomes
\be\label{Xddotbeta} 
\ddot{\bf X}_m = -\frac{1}{m}\gamma_m^{-3} (\mu_3-2\mu\cos\beta\gamma_w) \,, 
\ee 
so an angle smaller than\footnote{Notice the difference in the placement 
of Lorentz factors between this critical angle and that of
  Eq.~(\ref{xconstr}). This can be attributed to the additional
  transfer of momentum which can be thought of as an effective
  rescaling of the tension from $\mu\gamma^{-1}$ to $\mu\gamma$.}
$\beta_{\rm crit}=\arccos(\mu_3/2\mu\gamma_w)$ will result in positive
acceleration, which tends to increase the angle towards $\beta_{\rm
  crit}$. We may then expect the junction equilibrium to be stable. In
this section we will study this question quantitatively and for more
general string configurations, solving the equations of motion
numerically.

Perturbations of Y-junction configurations have been studied in
Ref.~\cite{Copeland:2006if} where the authors considered an initially
static three-string solution with $\dot s_i=0$ and studied propagation
of waves, transmitted and reflected at the junction, leading to
oscillatory behaviour for $s_i(t)$. Here, we will instead consider
deformations of basic solutions in which the effective tensions at the
junction are not balanced and the massive junction feels a non-trivial
force.

\subsection{Equal tensions $\mu_1=\mu_2$}

We start by considering the simplest case we have discussed above,
where the colliding strings have equal tension $\mu_1=\mu_2\equiv\mu$
and the zipper stays on the $x$-axis.  We first construct the
unperturbed solution on which our analysis will be based on.

\bigskip
\noindent{\it Unperturbed Solution}
\smallskip

We construct the unperturbed solution along the lines sketched in
Section~\ref{review}.  The idea is to write the solution along each of
the segments $JK_1$ and $JK_2$ (see Fig.~\ref{fig:unzipsaffin} below)
in terms of a unit direction vector and a transverse velocity, as in
Eq.~(\ref{wdparam}), and then determine ${\bf d}_i$ and ${\bf w}_i$
from the geometry.  Note that for $v\ne 0$ both $\bd$ and $\bw$ also
have finite $z$-components, which are not shown in perspective in
Fig.~\ref{fig:unzipsaffin} (see also Fig.~\ref{Junction3D}, {\it
  Left}).

\begin{figure}[h]
\centering \includegraphics[scale=0.7]{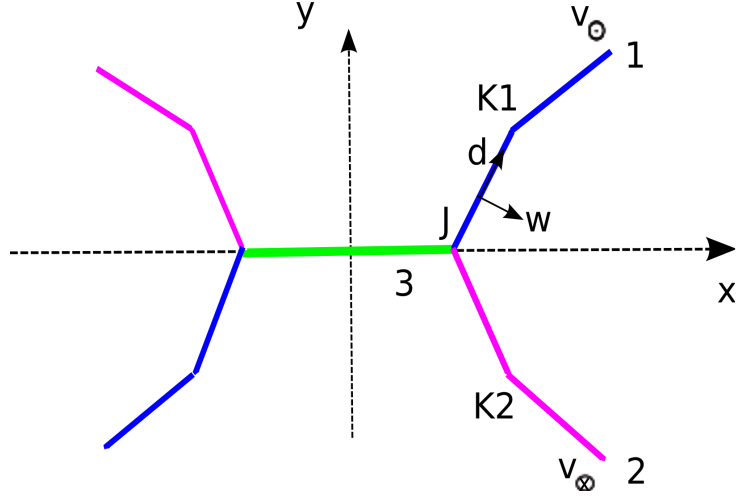}
\caption{Straight strings collision.}
\label{fig:unzipsaffin}
\end{figure}

The complete solution (for the $x>0$ half $x$-axis) in the conformal
gauge is: \be\label{soleqtens}
\begin{array}{lclcl}
\bx_1 &=& \left\{ \begin{array}{lcl} (-\sigma \gamma_v^{-1} \cos
  \alpha, - \sigma \gamma_v^{-1} \sin\alpha, + v t) \equiv {\bf
    x}_{1\infty} \\ - \sigma \gamma_w^{-1} {\bf d}_1 + {\bf w}_1 t
  \equiv {\bf x}_{1\rm f}
                           \end{array}
                  \right. 
                 &\begin{array}{rcl}
                     , \\
                     ,                
                   \end{array}&
                 \begin{array}{rcl}
                     \sigma &\le& \! -t  \\
                     -t &\le& \sigma \le s_1(t)                       
                   \end{array}
                  \nonumber\\ \\ \bx_2 &=& \left\{ \begin{array}{lcl}
                    (-\sigma \gamma_v^{-1} \cos \alpha, + \sigma
                    \gamma_v^{-1} \sin\alpha, - v t) \equiv {\bf
                      x}_{2\infty} \\ -\sigma \gamma_w^{-1} {\bf d}_2
                    + {\bf w}_2 t \equiv {\bf x}_{2\rm f}
                                               \end{array}
                  \right.
                  &\begin{array}{rcl}
                     , \\
                     ,                
                  \end{array}& 
                 \begin{array}{rcl}
                     \sigma &\le& \! -t \\
                     -t &\le& \sigma \le s_2(t)  
                   \end{array}
                  \\ \\ \bx_3 &=& (\sigma, 0, 0) &,& \;\;\; 0 \; \le
                  \, \sigma \, \le \, s_3(t) \nonumber
\end{array}
\ee 
Let us now determine ${\bf d}_i$, ${\bf w}_i$. First note that if
${\bf d}_1=(d_x,d_y,d_z)$ and ${\bf w}_1=(w_x,w_y,w_z)$, then symmetry
implies ${\bf d}_2=(d_x,-d_y,-d_z)$ and ${\bf w}_2=(w_x,-w_y,-w_z)$,
so we only need to consider ${\bf x}_1$ and ${\bf x}_3$.  Continuity
at the position of the junction ${\bf r}_J$ requires ${\bf
  x}_1(s_1(t),t)={\bf x}_3(s_3(t),t)$ so that
\be
\label{eq:rJ}
\br_J = \bw t -\frac{\sigma_J}{\gamma_w}\bd=(\dot{s}_3,0,0)t\,,  
\ee 
with $\sigma_J=s_1(t)=\dot s_1 t$.  Similarly, continuity at the kink implies 
${\bf x}_{1f}(-t,t)={\bf x}_{1\infty}(-t,t)$, that is
\be
\label{eq:rK1}
\br_{K_1}=\bw t -\frac{\sigma_{K_1}}{\gamma_w}\bd = \left(\frac{\cos
  \alpha}{\gamma_v}, \frac{\sin \alpha}{\gamma_v},v\right)t, 
\ee 
with $\sigma_{K_1}=-t$.  Since we are working in the conformal gauge 
($\bd\cdot\bw=0$, $\bd^2=1$) the invariant length of the segment 
$\vec{JK_1}={\bf r}_{K1}-{\bf r}_J$ is just $|\sigma_{K_1}-\sigma_J|=(1+\dot{s}_1)t$.  
From energy conservation (cf. Eq. (\ref{energycons}) with $\dot \gamma_m=0$) 
we have
\be
\label{encon}
\dot{s}_1=-\frac{\mu_3}{2\mu}\dot{s}_3\equiv-R\,\dot{s}_3\,, 
\ee 
so we can express $\bd$ and $\bw$ in terms of the original angle $\alpha$, $\gamma_v$,
$\dot s_3$ and the tension ratio $R=\mu_3/2\mu<1$ (but note these four quantities are 
related through Eq. (\ref{constr})).  Multiplying Eq.~(\ref{eq:rK1}) 
by $\dot{s}_1$ and adding it to Eq.~(\ref{eq:rJ}) we find
\be (1-R\dot{s}_3)\bw=(\dot{s}_3-R
\dot{s}_3\gamma_v^{-1}\cos\alpha, 
-R\dot{s}_3\gamma_v^{-1}\sin \alpha, 
-R\dot{s}_3v)~,  
\ee 
from which we read directly the components of $\bw$: 
\bea
w_x&=&\frac{(1-R\gamma_v^{-1}\cos \alpha)}{1-R\dot{s}_3}\dot{s}_3~,
\\ w_y&=&-\frac{R\gamma_v^{-1}\sin \alpha}{1-R\dot{s}_3}\dot{s}_3~,
\\ w_z&=&-\frac{Rv}{1-R\dot{s}_3}\dot{s}_3\,.  \eea 
The corresponding Lorentz factor is
\be\label{gammaw} \gamma_w=\frac{1-R\dot{s}_3}{\sqrt{A}}~, \ee
with 
\be
A=1-\dot{s}^2_3-2R\dot{s}_3(1-\gamma_v^{-1}\cos \alpha \dot{s}_3)\,.
\ee
Since $1-R\dot{s}_3$ is the invariant length of the segment $JK_1$ in
units of $t$, from Eq.~(\ref{gammaw}) it follows immediately that the
quantity $\sqrt{A}$ is the physical length of $JK_1$, in units of
$t$. Equivalence with $|\vec{JK_1}|=|\vec{{\bf x}}_1(-t,t)-{\bf
  x}_3(s_3(t),t)|=t\sqrt{1+\dot s_3[\dot s_3-2\cos{\alpha}/\gamma_v]}$
can be easily checked using Eq.~(\ref{constr}). Finally, for the
components of $\bd$ we find
\bea 
\label{dx}
d_x&=&-\frac{(\gamma_v^{-1}\cos \alpha-\dot{s}_3)}{\sqrt{A}}~,
\\ d_y&=&-\frac{\gamma_v^{-1}\sin \alpha}{\sqrt{A}}~,
\\ d_z&=&-\frac{v}{\sqrt{A}} \,. \eea
The signs are in agreement with our convention that $\sigma$ is
negative and increasing towards the junction on ${\bx}_1$, so that
$d_x$ is identified with $-\cos\beta$ in
Eq.~(\ref{cosbetavnonzero}). The angle $\phi$, describing the rotation
of the plane spanned by $\bx_{1\rm f}$ and $\bx_{2\rm f}$ with the
$x$-axis, is given by
\be\label{tanphi}
\tan\phi=d_z/d_y=v\gamma_v\csc{\alpha} \,.  \ee
In terms of the angles $\beta$ and $\phi$ the solution between the
junction and the kink, $-t \le\sigma\le s_1(t)$, can be written as
\be\label{stringsatbetav} \bx_{1\rm f} = (-\sigma \gamma_w^{-1}
\cos\beta + w \sin\beta\, t, (- \sigma \gamma_w^{-1} \sin\beta -
w\cos\beta\, t)\cos\phi, (- \sigma \gamma_w^{-1} \sin\beta -
w\cos\beta\, t)\sin\phi)~, \ee
and similarly for $\bx_{2\rm f}$. This can now be compared directly to the
solution (\ref{stringsatbeta}) which corresponds to $v=0\Rightarrow
\phi=0$ .

We will next introduce a mass on the junction which will allow us to
deform this solution by breaking the effective tension balance at the
expense of having non trivial dynamics for the junction $\ddot
\bX_m\ne 0$. 

\bigskip
\noindent{\it Massive Junction and Deformed Solution}
\smallskip

Keeping the enhanced symmetry of the problem for $\mu_1=\mu_2$, we
look for a more general ansatz for $\bx_{1\rm f}$ (equivalently
$\bx_{2\rm f}$) in Eq.~(\ref{soleqtens}), which will allow for
non-trivial acceleration at $\br_J=\bx_{1\rm f}(s_1(t),t)=\bX_m$.
Since the junction can now have $\dot \gamma_m\ne 0$ we cannot assume
$\dot s_1=-R\dot s_3$, but instead $s_1(t)$ and $s_3(t)$ (we still
have $\dot s_2=\dot s_1$) must satisfy the evolution
(\ref{ddotXm2}) and the constraint equation (\ref{energycons}).  Note
that the kink is still moving with the speed of light along the
original strings.
\\
Let us write the equation for the segment $JK_1$:
\be
\bx_1=\mathbf{W}(t)-\frac{\sigma}{\gamma_w}\bd(t)~,
\ee
and for the junction $J$:
\be
\br_J=\mathbf{W}(t)-\frac{s_1(t)}{\gamma_w}\bd(t)=(s_3(t),0,0)~,
\ee 
while for the kink $K_1$:
\be
\br_{K_1}=\mathbf{W}(t)+\frac{t}{\gamma_w}\bd(t)=\left(\frac{\cos
  \alpha}{\gamma_v},\frac{\sin \alpha}{\gamma_v},v\right)t~.  \ee
Hence, subtracting we get
\be
\label{eq:unitvec}
[t+s_1(t)]\bd(t)=\gamma_w\left(\frac{\cos
  \alpha}{\gamma_v}t-s_3(t),\frac{\sin \alpha}{\gamma_v}t,vt\right),
\ee
from which, after squaring and using $\bd^2=1$, we obtain
\be
t+s_1(t)=\gamma_w\sqrt{t^2+[s_3(t)]^2-2(\cos \alpha/\gamma_v)ts_3(t)}~.
\ee
Energy conservation, Eq.~(\ref{energycons}), leads to
\be
\label{eq:en}
2\mu \dot{s}_1+\mu_3\dot{s}_3+m(1-\dot{s}^2_3)^{-3/2}\dot{s}_3\ddot{s}_3=0~,
\ee
and from Eq.~(\ref{ddotXm2}) we get
\be
\label{eq:s3}
\ddot{s}_3=-\frac{1}{m}(1-\dot{s}^2_3)^{3/2}[\mu_3-2\mu \gamma_w(t) d_x(t)]~.
\ee
We can now solve numerically for $s_3(t)$.  For initial conditions
corresponding to tension and momentum transfer balance,
Eq. (\ref{Xddotbeta}) with $\ddot \bX_m=0$ ($\beta=\beta_{\rm crit}$),
the solution is identical to the one considered above. Let us now
consider a deformation of the initial configuration with $\ddot
\bX_m\ne 0$, $\beta\ne \beta_{\rm crit}$, and study its evolution. In
this case, we find that $\dot s_3(t)$ oscillates around the CKS value
$(2\mu \gamma_v^{-1} \cos\alpha -
\mu_3)/(2\mu-\mu_3\gamma_v^{-1}\cos\alpha)$ with decaying
amplitude. Hence, as we may have expected from the discussion following
Eq.~(\ref{Xddotbeta}), the local angle $\beta$ exhibits damped
oscillations around $\beta_{\rm crit}$.  This is shown in
Fig.~\ref{fig:s3mass}.
\begin{figure}[h]
\centering \includegraphics[scale=0.95]{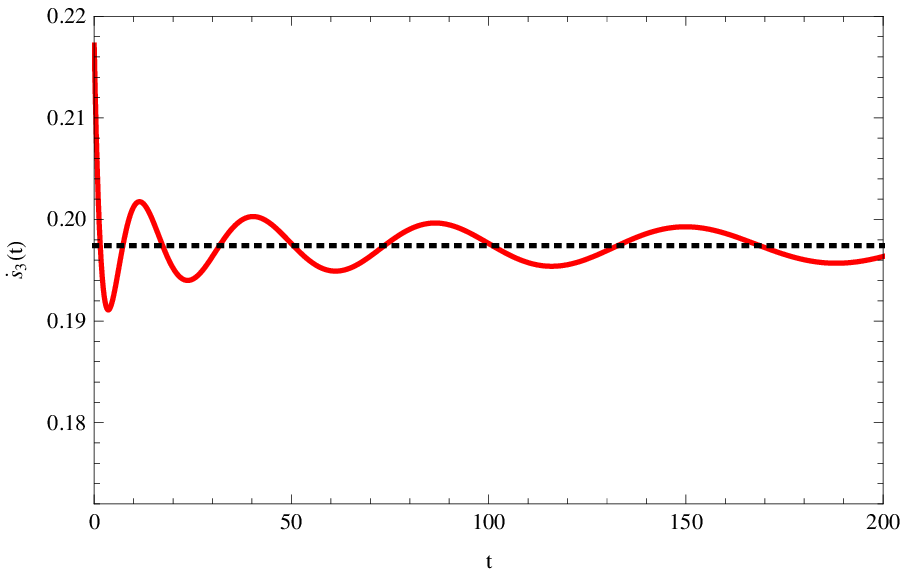}
\includegraphics[scale=0.45]{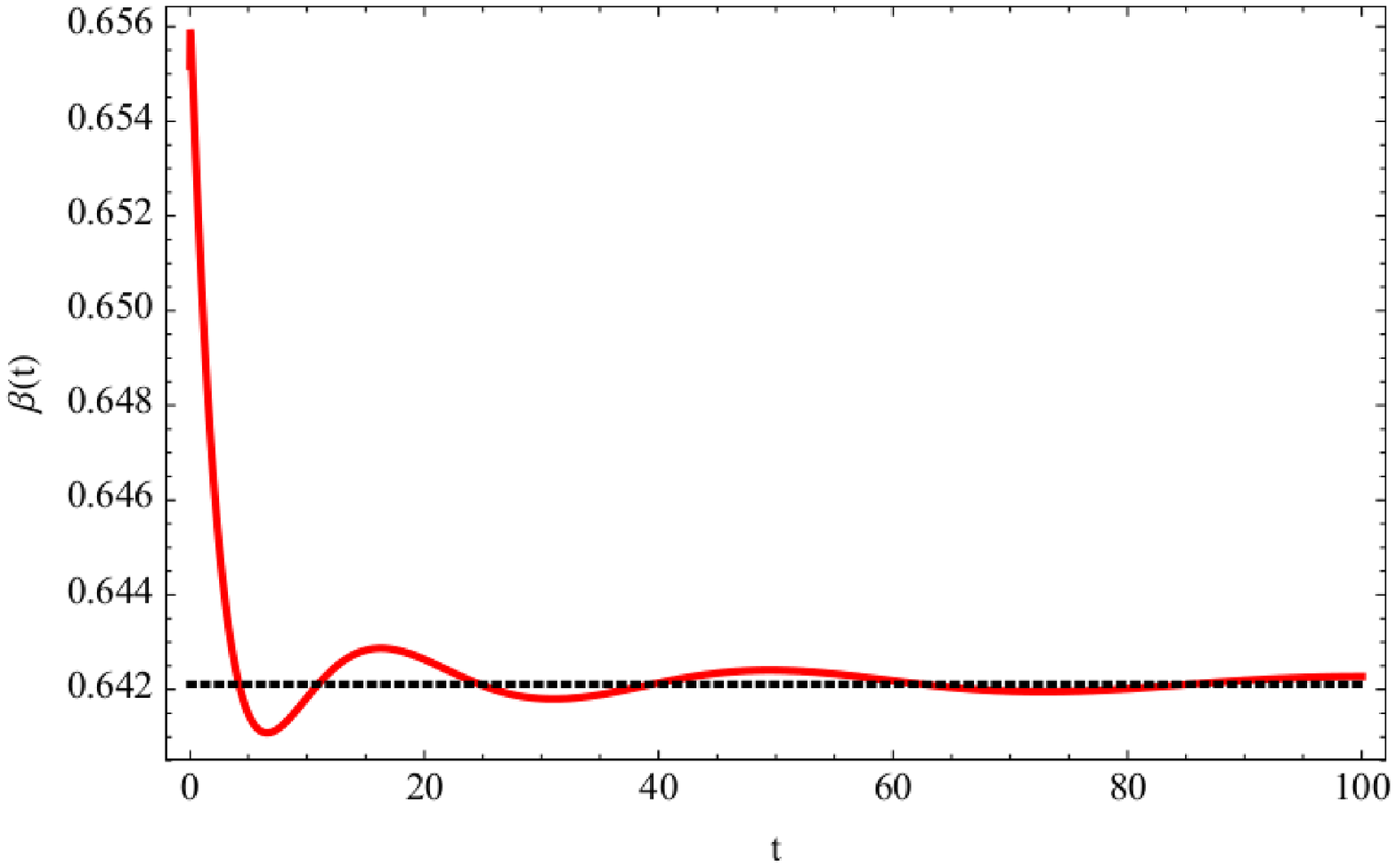}
\caption{{\it Left:} The evolution of $\dot s_3$ for $\mu=1$,
  $\mu_3=\sqrt{2}\mu$, $m=\sqrt{\mu}$, $\alpha=\pi/6$ and $v=0.4$,
  shown together with the unperturbed case $\dot{s}_3=\rm{const.}$,
  given by the CKS solution. {\it Right:} Evolution of the angle
  $\beta(t)$ for the same parameters. The dotted line shows the
  critical angle for tension and momentum transfer balance.}
\label{fig:s3mass}
\end{figure}

Therefore, even though the constant zipper growth prediction, $\dot
s_3=\rm const$, was based on special string configurations satisfying
the balance condition (\ref{junctionbalance1}), these configurations
are stable and deformations of the balance condition lead to identical
asymptotic behaviour. In conclusion, unzipping cannot occur by simply
destabilising the angle $\beta_{\rm crit}$.

\subsection{Unequal tensions $\mu_1\ne \mu_2$}
\label{sec:unzipgeom}

Let us consider the general case where the colliding strings have 
unequal tensions $\mu_1\ne \mu_2$. For a massless junction, this has
been studied in Refs.~\cite{Copeland:2006if, Copeland:2007nv,
  Bevis:2008hg}.  The geometry of the problem involves two extra
parameters, the angle of the zipper with the $x$-axis and the zipper
velocity $u$ along the $z$-axis (see Fig.~\ref{fig:configuneq}), which
are found by solving~\cite{Copeland:2006if}
\be 0=u^4 S^2 \sin^2
\alpha + u^2 [R^2(1-v^2)+S^2(v^2\cos^2 \alpha-\sin^2 \alpha)] -S^2 v^2
\cos^2 \alpha \ee and \be \tan \theta = \frac{u}{v}\tan \alpha~, \ee
where $R=\mu_3/(\mu_1+\mu_2)$ and $S=(\mu_1-\mu_2)/(\mu_1+\mu_2)$.
Note that due to the asymmetry of the configuration, we no longer have
$s_1(t)=s_2(t)$.
\begin{figure}[h]
\centering \includegraphics[scale=0.7]{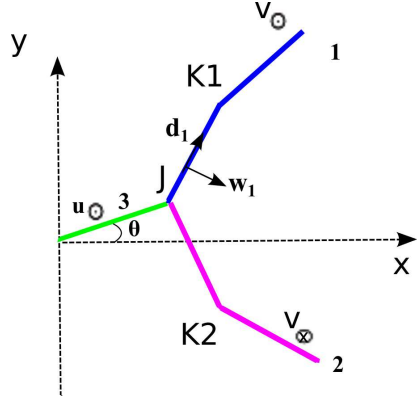}
\caption{Schematic representation of the three-string configuration in the general case where the colliding strings have unequal tensions $\mu_1 \neq \mu_2$. Note we only show the $x>0$ part of the configuration.}
\label{fig:configuneq}
\end{figure}
The general ansatz for the segment $JK_1$ is
\be
\bx_1=\mathbf{W}_1(t)-\frac{\sigma}{\gamma_{w_1}}\bd_1(t)~, \ee 
while
for $JK_2$ it reads
\be
\bx_2=\mathbf{W}_2(t)-\frac{\sigma}{\gamma_{w_2}}\bd_2(t)~.  \ee 
For
the junction $J$ we have 
\be
\br_J=\mathbf{W}_1(t)-\frac{s_1(t)}{\gamma_{w_1}}\bd_1(t)=(\gamma^{-1}_u
s_3(t) \cos \theta,\gamma^{-1}_u s_3(t) \sin \theta,ut)~, \ee 
while
for the kink $K_1$:
\be
\br_{K_1}=\mathbf{W}_1(t)+\frac{t}{\gamma_{w_1}}\bd_1(t)=\left(\frac{\cos
  \alpha}{\gamma_v},\frac{\sin \alpha}{\gamma_v},v\right)t~.  \ee
Following the same procedure as before, we find 
\be [t+s_1(t)]\bd_1(t)=\gamma_{w_1}(\gamma^{-1}_v \cos \alpha \, t -
\gamma^{-1}_us_3(t)\cos \theta, \gamma^{-1}_v \sin \alpha t -
\gamma^{-1}_u s_3(t) \sin \theta, vt-ut) \ee and \be t+s_1(t) =
\gamma_{w_1}
\sqrt{t^2(1+u^2-2uv)+\gamma^{-2}_u[s_3(t)]^2-2ts_3(t)\gamma^{-1}_u
\gamma^{-1}_v\cos
  (\alpha-\theta)}.  \ee
Note that one can easily write down the analogous equations for
$\bd_2(t)$ and $\gamma_{w_2}$.  From Eq.~(\ref{ddotXm2}) we get 
\be
\gamma^{-1}_u\ddot{s}_3 \cos \theta=-\frac{1}{m}(1-\dot{s}^2_3
\gamma^{-2}_u - u^2)^{3/2} [\mu_3 \gamma_u \cos \theta -\mu_1
  \gamma_{w_1}(t) d_{x_1}(t)-\mu_2 \gamma_{w_2}(t) d_{x_2}(t)]~.  \ee
The above equation is numerically solved to find the evolution of
$s_3(t)$, together with Eq.~(\ref{dotseqv1}) for $s_1(t)$ and
$s_2(t)$. Taking initial conditions corresponding to tension and
momentum transfer balance, as before, the solution is identical to the
massless case. Considering a deformation of the initial configuration
we again find that $\dot s_3(t)$ oscillates around the CKS value with
decaying amplitude, as expected from the results of the symmetric
case. This is shown in Fig.~\ref{fig:s3massasym}.
\begin{figure}[h]
\centering
\includegraphics[scale=0.95]{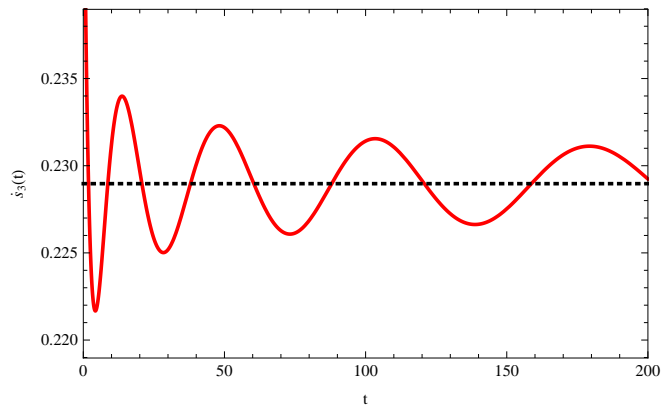}
\caption{The evolution of $\dot s_3$ for $\mu_1=1$, $\mu_2=0.7$, $\mu_3=1.2$, $m=\sqrt{\mu_1}$, $\alpha=\pi/6$ and $v=0.4$, 
shown together with the unperturbed case $\dot{s}_3=\rm{const.}$ given by the CKS solution.}
\label{fig:s3massasym}
\end{figure}

\section{Unzipping Mechanisms}
\label{unzipping}
In this section we investigate whether monopole or string forces, and string curvature for loops with junctions, are viable unzipping mechanisms.

\subsection{Monopole Forces}

In the previous sections we saw that unzipping cannot occur by simply perturbing the angle $\beta_{\rm crit}$. 
The dynamics of free junctions is such that perturbations get damped and the configuration stabilises at the critical 
angle. This could of course change if the junction was subject to external forces. We now move on to allow for such
forces exerted on the junction and ask whether this can lead to non-trivial evolution of string-junction 
configurations. We are interested in solutions exhibiting accelerated zipping and, more interestingly, 
decelerated zipping possibly leading to unzipping.

Physically, junction forces can arise, for example, in hybrid string-monopole networks, where the monopoles 
are subject to long-range interactions. In this context, we consider the junction of our three-string configuration 
as a monopole of mass $m$, and introduce a force due to another monopole (or anti-monopole) located at 
some distance apart. For local monopoles, the force is just electromagnetic (Coulomb) interaction due to their 
magnetic charge, and for global monopoles the force is independent of distance. One can also consider the drag 
force felt by the monopoles due to their interaction with charged particles in a plasma.         

We will start with the action (\ref{action}), having both string world-volume and monopole world-line 
contributions, but we will now introduce an additional world-line piece, giving rise to a force term for the 
monopole at the junction. The force can be conveniently described through the integral of a 1-form, 
say $A$, on the world-line. $A$ is to be thought of as a background field (a 1-form gauge potential) 
which is pulled-pack on the world-line, yielding the following reparametrisation-invariant piece:
\be\label{Sp}
S_{\rm monopole} = - q \int A = - q \int d\tau  A_\nu \dot{X}^\nu_m \,,
\ee     
with $A_\nu$ evaluated on the monopole world-line $X^\mu_m(\tau)$.  This is the standard coupling 
of a relativistic particle with electric charge $q$ to the electromagnetic gauge potential, giving rise to 
the Lorentz force $q({\bf E} + {\bf v \times \bf B})$ in standard notation classical electrodynamics. 
Equivalently, it also describes the coupling of a monopole of magnetic charge $q$ to the `magnetoelectric' 
potential, giving rise to the corresponding Lorentz force $q({\bf B} - {\bf v \times \bf E})$.  Here, 
we will use it as a convenient phenomenological action to construct a monopole force and study its 
effect on junction dynamics. 

The total action is 
\bea \nonumber
\label{action2}
S&=&-\displaystyle\sum_i      \mu_i      \int     d\tau      d\sigma_i
\Theta(s_i(\tau)-\sigma_i)     \sqrt{-x'^2_i\dot{x}^2_i}     \\   \nonumber  &&+
\displaystyle\sum_i       \int        d\tau       f_{i\mu}       \cdot
                 [x^\mu_i(\tau,s_i(\tau))-X^\mu_m(\tau)]-m\int    d\tau
                 \sqrt{\dot{X}^2_m} \\
                 &&- q \int d\tau  A_\nu(X^\mu_m(\tau)) \dot{X}^\nu_m ~.  \eea
Varying this action with respect to $x^\mu_i$, the terms proportional to $\Theta(s_i(\tau)-\sigma)$ give the wave equation 
\be
\ddot{x}^\mu_i-x^{\mu\prime\prime}_{i}=0\,,  
\ee
and the boundary terms proportional to $\delta(s_i(t)-\sigma)$ give 
\be
\mu_i (x^{\mu \prime}_i + \dot{s}_i \dot{x}^\mu_i)=-f^\mu_i
\ee 
at the junction, as before. Varying the Lagrange multipliers $f^\mu_i$ we obtain the boundary condition
\be
X^\mu_m(\tau)=x^\mu_i(\tau,s_i(\tau)),
\ee while varying $X^\mu_m$ we get
\be
m\frac{d}{d\tau}\left(\frac{\dot{X}^\mu_m}{\sqrt{\dot{X}^2_m}}\right) = \displaystyle\sum_i f^\mu_i
+qF^{\mu \nu} (\dot{X}_m)_\nu,
\ee where $F_{\mu\nu}= \partial_\mu A_\nu - \partial_\nu A_\mu$.
We can therefore write
\be
\label{eq:Lorentz}
m\frac{d}{d\tau}\left(\frac{\dot{X}^\mu_m}{\sqrt{\dot{X}^2_m}}\right) = -\displaystyle\sum_i \mu_i(x^{\mu \prime}_i +  \dot{s}_i \dot{x}^\mu_i) +qF^{\mu \nu} (\dot{X}_m)_\nu. 
\ee
We now have an extra term in the equations of motion for the vertex, which we will use to model the contribution 
of monopole forces at the junction. Writing the 4-potential in terms of a scalar potential and 3-vector potential, 
$A^\mu = (\phi, {\bf A})$, and defining
\be
{\bf \mathcal{E}} = -\dot{\bf A} - \nabla \phi \,, \;\;\; {\bf {\mathcal B}} = \nabla \times {\bf A},
\ee
the energy conservation equation (i.e. the $\mu=0$ component of  Eq.~(\ref{eq:Lorentz})) becomes
\be
\label{eq:enconsLorentz}
m\dot{\gamma}_m + \sum_i \mu_i \dot{s}_i = q\, {\bf \mathcal{E}} \cdot \dot{\bX}_m,
\ee while the $i$-th components give
\be
m\frac{d}{dt}(\gamma_m \dot{\bX}_m)= -\sum_i \mu_i(\bx^\prime_i+\dot{s}_i\dot{\bx}_i)
+q({\bf \mathcal{E}} + \dot{\bX}_m \times {\bf \mathcal{B}}).
\ee
After a little algebra using Eqs.~(\ref{gauge}) and (\ref{dotbXm}) as well as Eq.~(\ref{eq:enconsLorentz}), we find
\be
\label{ddotXm3}
\ddot{\bX}_m=-\frac{1}{m}\gamma^{-3}_m\displaystyle\sum_i\mu_i\frac{\bx^\prime_i}{\bx^{\prime
    2}_i}-\frac{q}{m}\gamma^{-1}_m\, \left[\gamma^{-2}_m {\mathcal{E}}_{\parallel}  + {\mathcal{E}}_{\perp} + \dot{\bX}_m \times {\bf \mathcal{B}}\right]\,,
\ee
where $\mathcal{E}_{\parallel}, \mathcal{E}_{\perp}$ are the components of $\bf\mathcal{E}$ 
parallel and transverse to $\dot{\bX}_m$ respectively.
    
With $A$ the standard electromagnetic gauge 4-potential, we have ${\bf \mathcal{E}}={\bf E}$ and ${\bf \mathcal{B}}={\bf B}$ 
in standard notation, so the rightmost quantity on the right hand side of Eq. (\ref{ddotXm3}) is the Lorentz force for an 
electrically charged particle, as we expect from the above discussion.  Similarly, if $A$ is the magnetoelectric 4-potential 
then ${\bf \mathcal{E}}={\bf B}$ and ${\bf \mathcal{B}}=-{\bf E}$ in standard notation, giving rise to the Lorentz force 
for a magnetically charged particle. As mentioned above, here we will use $A$ as a phenomenological potential to construct 
a monopole force of the desired type. Let us for example consider a constant force, as is the case for global monopoles, and 
take it for simplicity to be aligned with string 3 (i.e. the zipper along the $x$-axis). Then $\mathcal{E}_{\parallel}=\mathcal{E}$, 
$\mathcal{E}_{\perp}=0$ in (\ref{ddotXm3}), and in the equal tension case the junction only moves along the $x$-axis. 
We choose ${\bf\mathcal{B}}=0$ and ${\bf \mathcal{E}}$ a constant vector tangent to string $3$, such that
\be
{\bf \mathcal{E}} = \epsilon \, \hat{x}\,,
\ee 
with $\epsilon$ a constant. We can then write the $x$-component of Eq.~(\ref{ddotXm3}) as
\be
\ddot{s}_3=-\frac{1}{m}(1-\dot{s}^2_3)^{3/2}[\mu_3 + q\epsilon -2\mu \gamma_w(t) d_x(t)]~,
\ee and the energy conservation equation (\ref{eq:enconsLorentz}) as
\be
m\dot{\gamma}_m + \sum_i \mu_i \dot{s}_i = q \epsilon \dot{s}_3~.
\ee
The last two equations generalise Eqs. (\ref{eq:s3}) and (\ref{energycons}) by including a constant force along the zipper.  
We now investigate numerically the effect of this additional force. The initial conditions we use are the ones corresponding to tension and momentum transfer balance, i.e. in our initial configuration the angle $\beta$ has the critical value satisfying Eq. (\ref{Xddotbeta}) for $\ddot{\bf X}_m=0$. 
The force is felt for $t>0$, and, for positive $\epsilon$, its effect is to cause unzipping of the initial configuration over a timescale determined by the mass of the junction $m$ and the magnitude of the force $q\epsilon$. In Fig.~\ref{fig:s3pert} we show the evolution of $s_3(t)$ for the three-string symmetric configuration studied before, but we have now included a monopole-like force term with a fixed magnitude $q\epsilon=0.5$. Having fixed $q\epsilon$, the unzipping timescale is controlled by the monopole mass. For lower mass the unzipping starts earlier and takes shorter to complete, while for larger masses, unzipping starts later and completes over a longer period. In the figure we show examples for $m=\sqrt{\mu}/2$ (top left), $m=\sqrt{\mu}/4$ (top right), $m=2\sqrt{\mu}$ (bottom). The $q\epsilon=0$ case, with $\dot s_3(t)={\rm const.}$ solution, is shown by the dashed black lines.   
\begin{figure}[H]
\centering \includegraphics[scale=0.55]{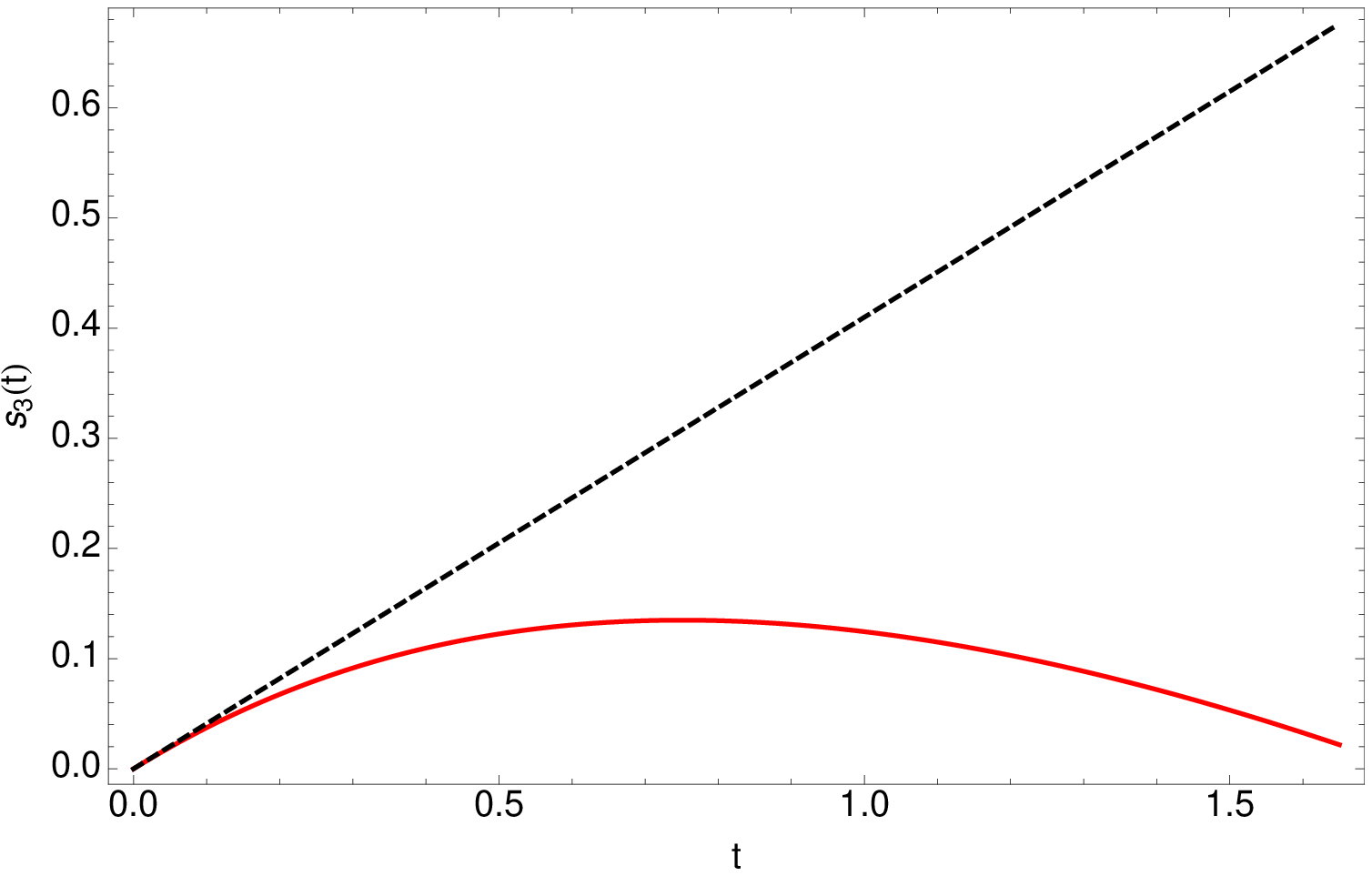}
\includegraphics[scale=0.55]{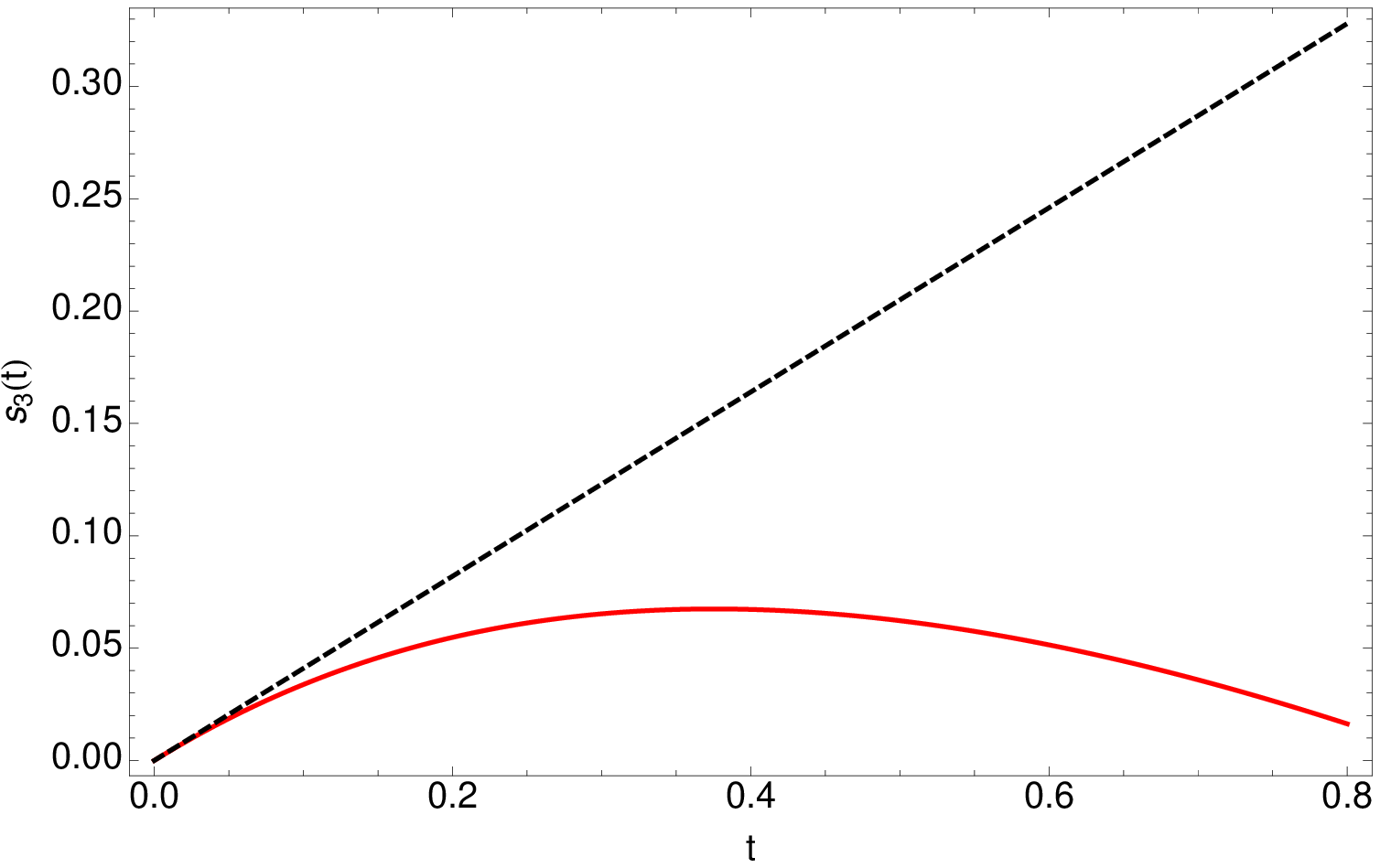}
\includegraphics[scale=0.55]{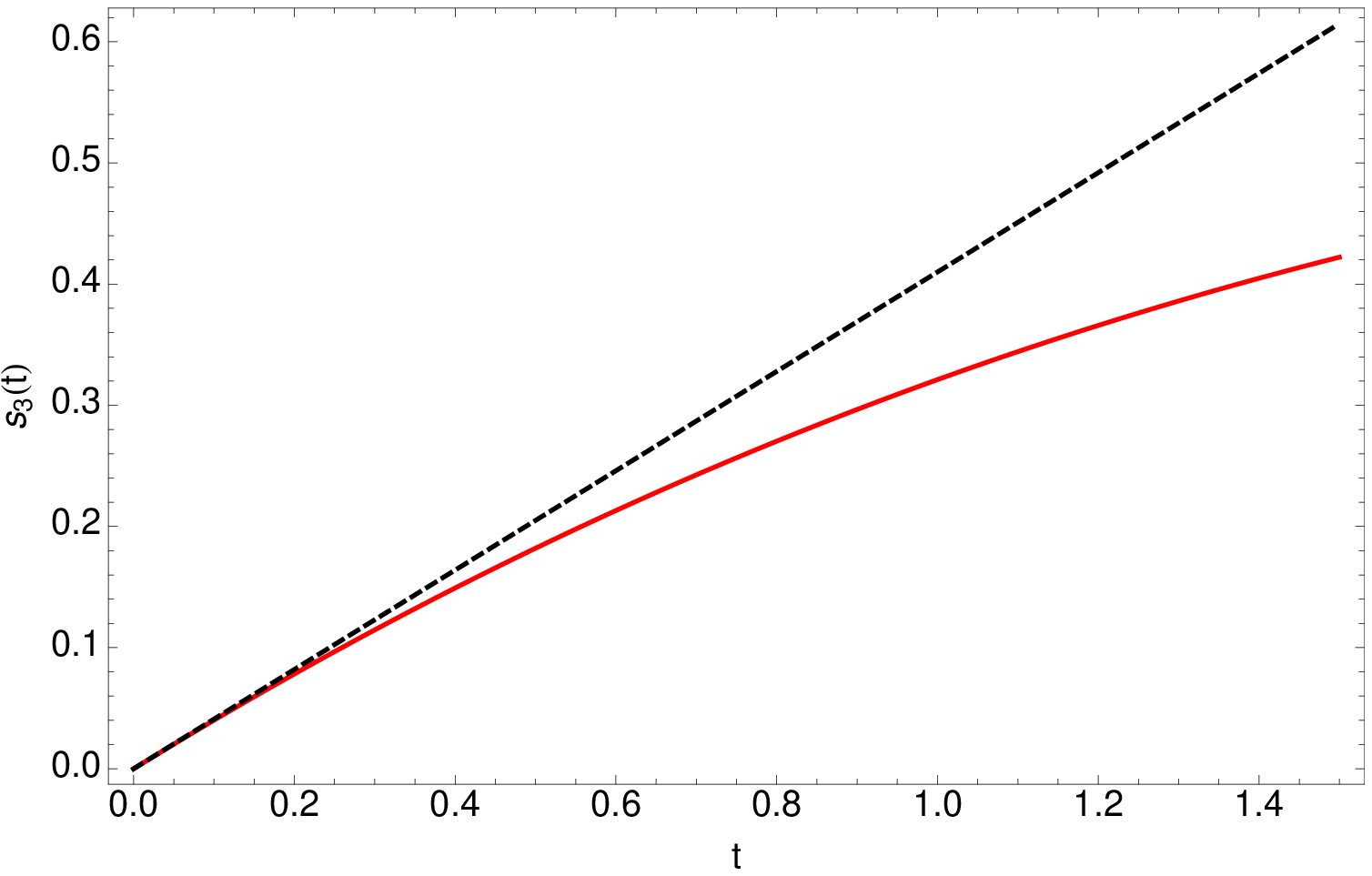}
\caption{The evolution of $s_3$ including an extra force towards the junction with magnitude $q\epsilon=0.5$ (solid red lines), for $\mu=1$,
  $\mu_3=\sqrt{2}\mu$, $\alpha=\pi/6$ and $v=0$, for $m=\sqrt{\mu}/2$ (top left), $m=\sqrt{\mu}/4$ (top right), $m=2\sqrt{\mu}$ (bottom), shown together with the unperturbed case given by the CKS solution (dashed black lines). }
\label{fig:s3pert}
\end{figure}

The situation is analogous for a force also having components transverse to the zipper. The projection of
the force along the zipper produces an acceleration for $s_3(t)$, as above. From Eq.~(\ref{ddotXm3}) it is 
clear that allowing for a non-trivial component of the monopole force in the direction transverse to the 
zipper, $\mathcal{E}_\perp\ne 0$, breaks the symmetry of the problem leading to a situation analogous to 
the case $\mu_1\ne \mu_2$.  

\subsection{String Forces}

Let us move on to consider the effect of string forces on the evolution of junctions. Such forces 
can arise in a wide range of physically relevant situations. Examples include long range interactions between 
field theory solitons (e.g. global string interactions), friction due to particle scattering in a plasma, exchange 
of dilatons for cosmic superstrings, and D-brane forces arising from fundamental strings stretching between 
two branes. In analogy to our approach for modelling monopole forces in the previous subsection, we will 
phenomenologically describe string forces through a background field. 

A string force can be readily described by the integral of a background 2-form field pulled-back on the string 
worldsheet. For each string, $i$, we introduce the following reparametrisation-invariant worldsheet contribution:
\be\label{Ss}
S_{\rm string} = q_i \int B = q_i \int d\tau d\sigma_i B_{\mu\nu} \epsilon^{\alpha\beta} \partial_\alpha x^\mu_i 
\partial_\beta x^\nu_i \,,
\ee     
with $B_{\mu\nu}$ evaluated on the $i$-string world-sheet $x^\lambda(\tau,\sigma_i)$ and $q_i$ 
constants. This is, for example, how the fundamental string couples to the Neveu-Schwarz 2-form 
and the D1-brane to the Ramond-Ramond 2-form, giving rise to a stringy interaction analogous 
to electromagnetism. Here, we will use this action as a phenomenological tool to construct a 
convenient string force.            

The full action reads 
\bea \nonumber
\label{action3}
S&=&-\displaystyle\sum_i      \mu_i      \int     d\tau      d\sigma_i
\,\Theta(s_i(\tau)-\sigma_i)\,     \sqrt{-x'^2_i\dot{x}^2_i}     \\   \nonumber  &&+
\displaystyle\sum_i       \int        d\tau       f_{i\mu}       \cdot
                 [x^\mu_i(\tau,s_i(\tau))-X^\mu_m(\tau)]-m\int    d\tau
                 \sqrt{\dot{X}^2_m} \\
                 &&+ \displaystyle\sum_i q_i \int d\tau d\sigma_i\, \Theta(s_i(\tau)-\sigma_i)\, B_{\mu\nu}(x^\lambda(\tau,\sigma_i)) 
                 \epsilon^{\alpha\beta} \partial_\alpha x^\mu_i \partial_\beta x^\nu_i\,,
\eea
where we have introduced a Heaviside $\Theta$-function in the last term accounting for the moving junction. 
Upon variation, this produces a $\delta$-function contribution localised at the junction so the boundary term in 
Eq. (\ref{boundaries}) acquires a term proportional to $q_i B^\mu_{\,\nu} (\dot x_i^\nu+x_i^{\prime\nu} \dot s_i(t))$. 
Consequently, Eq. (\ref{varXm}) receives a contribution proportional to 
$\sum\limits_i q_i B^\mu_{\,\nu} (\dot x_i^\nu+x_i^{\prime\nu} \dot s_i(t))$. 

The conformal temporal gauge we have adopted in the preceding sections is too constraining for the present case. 
For a non-trivial string force there is no residual freedom to impose the temporal condition on top of the conformal gauge 
-- doing so would violate the equations of motion. It is then convenient to keep the temporal ($x^0=\tau$) and 
transverse ($\dot{\bx}_i \cdot \bx^{\prime}_i=0$) conditions, but relax the tracelessness condition $\dot{\bx}^2_i+\bx^{\prime 2}_i=1$. 
We then define the scalar quantities 
\be\label{epsilon}
\varepsilon_i \equiv \frac{\sqrt{-\gamma_i}}{\dot x_i^2}=\sqrt{\frac{\bx_i'^2}{1- \dot \bx_i^2}}\,.
\ee
In the conformal temporal gauge these quantities are effectively set to unity, but in this transverse temporal gauge 
they are dynamical and their dynamics is governed by the 0-th component of the equations of motion. In this gauge 
the invariant length of the $i$-th string is given by $\int \varepsilon_i d\sigma$, while physical velocities remain transverse 
to the strings. 

The string equation of motion in the presence of our string force becomes
\be\label{stringforce}
\frac{\mu_i}{\sqrt{-\gamma_i}} \left[ \frac{\partial}{\partial\tau}(\varepsilon_i\dot{x}^\mu_i)-\frac{\partial}{\partial\sigma}
\left(\frac{x^{\mu\prime}_{i}}{\varepsilon_i}\right)\right]=q_i \epsilon^{\alpha\beta} \partial_\alpha x_i^\lambda 
\partial_\beta x_i^\nu H^\mu_{\lambda\nu}=q_i (\dot x_i^\lambda x_i^{\prime\nu} - x_i^{\prime\lambda} 
\dot x_i^\nu) H^\mu_{\lambda\nu}\,,  
\ee 
where $H_{\mu\nu\lambda}=3\,\partial_{[\mu} B_{\nu\lambda]}$, the square brackets denoting 
antisymmetrisation with respect to the enclosed indices. This splits into evolution equations for 
$\varepsilon_i$ and ${\bf x}_i$: 
\be
\left\{ \begin{array}{lcl} 
 \dot\varepsilon_i = \frac{q_i}{\mu_i} (\dot x_i^\lambda x_i^{\prime\nu} - x_i^{\prime\lambda} 
\dot x_i^\nu) H^0_{\lambda\nu} {\bf x}^2_i \varepsilon_i  \\ 
(\varepsilon_i \dot{\bf x}_i)\dot{} - \left(\frac{{\bf x}^{\prime}_{i}}{\varepsilon_i}\right)^{\prime} =  
\frac{q_i}{\mu_i} \sqrt{-\gamma_i} (\dot x_i^\lambda x_i^{\prime\nu} - x_i^{\prime\lambda} 
\dot x_i^\nu) {\bf H}_{\lambda\nu}
          \end{array}
 \right.\,,
\ee
where we have denoted the spatial components of $H^\mu_{\lambda\nu}$ as ${\bf H}_{\lambda\nu}$.
In this gauge the boundary term at the junction (cf. equation (\ref{boundaries})) reads
\be\label{Bndryforce}
\mu_i\left(\frac{x^{\mu\prime}_i}{\varepsilon_i}+\varepsilon_i \dot{s}_i\dot{x}^\mu_i\right) -2q_i B^\mu_{\,\nu} (\dot x_i^\nu+x_i^{\prime\nu} \dot s_i(t))=-f^\mu_i\,,  
\ee
leading to the following equation of motion for the massive junction (cf. equation (\ref{varXm})):
\be\label{eomXmforce}
m\frac{d}{d\tau}\left(\frac{\dot{X}^\mu_m}{\sqrt{\dot{X}^2_m}}\right)
=2 \displaystyle\sum_i q_i B^\mu_{\,\nu} (\dot x_i^\nu+x_i^{\prime\nu} \dot s_i(t))
-\displaystyle\sum_i\mu_i\left(\frac{x^{\mu\prime}_i}{\varepsilon_i}+\varepsilon_i\dot{s}_i\dot{x}^\mu_i\right)\,.
\ee
The toy model we would like to construct in the context of our 3-string Y-junction configuration in 
Fig.~\ref{Junction3D} ({\it Left}) is a ``Hook-Yukawa" force between strings 1 and 2.  In other words, we are 
taking the magnitude of the force to scale with the product of a linear factor and an exponentially 
damped factor in the $x^2\equiv y$ direction. This is a phenomenological choice motivated from the requirement 
that the force be localised near the junction. Close to the junction the force is taken to be linear 
(Hook-like), but it gets exponentially damped away from the junction, as the $y$-separation of the 
strings becomes large:
\be\label{HookYukawa} 
F=-k(y_1-y_2)e^{-M(y_1-y_2)}\,,
\ee    
with $k$ and $M$ positive constants. 

With our choice of gauge, for which the physical velocities are transverse to the string segments, 
it is convenient to construct the force transverse to the segments too, which is consistent with the 2-form 
nature of the force potential. 
Remembering that $x^0\!=\!\tau$ and setting $x^1\!\equiv\! x$, $x^2\!\equiv\! y$ and 
$x^3\!\equiv\! z$, the equations of motion (\ref{stringforce}) for string 1 in our simple planar configuration of 
Fig.~\ref{Junction3D} ({\it Left}) read
\bea\label{eomforce}
\dot\epsilon_1 &=& \frac{2q_1}{\mu_1}(\dot x_1 y_1^\prime - x_1^\prime \dot y_1) H^0_{12} (1-\dot x_1^2-\dot y_1^2) \epsilon_1~, \nonumber \\
\mu_1 \left[\ddot x_1 - \frac{1}{\epsilon_1^2} x_1^{\prime\prime} - \frac{1}{\epsilon_1}\left(\frac{1}{\epsilon_1}\right)^\prime x_1^\prime \right]  &=& 2 q_1 y^{\prime}_1 (1-\dot x_1^2-\dot y_1^2) H^1_{02} \equiv q_1 k (y_1-y_2) e^{-M(y_1-y_2)} (1-\dot x_1^2-\dot y_1^2) \sin\beta~,\nonumber \\
\mu_1 \left[\ddot y_1 - \frac{1}{\epsilon_1^2} y_1^{\prime\prime} - \frac{1}{\epsilon_1}\left(\frac{1}{\epsilon_1}\right)^\prime y_1^\prime \right] &=& 2 q_1 x^{\prime}_1 (1-\dot x_1^2-\dot y_1^2) H^2_{01} \equiv - q_1 k (y_1-y_2) e^{-M(y_1-y_2)} (1-\dot x_1^2-\dot y_1^2) \cos\beta~,  \nonumber \\  
z_1 &=& 0 \,, 
\eea   
where $\beta$ is the local string orientation, $\epsilon_1=\sqrt{(x_1^{\prime2}+y_1^{\prime2})/(1-\dot x_1^2-\dot y_1^2)}$, and we have used $H^0_{12}=H^1_{02}=-H^0_{21}$ in our conventions. We have chosen $H^0_{12}$ to correspond to the phenomenological force  (\ref{HookYukawa}), which can be generated by a static potential, $\partial_0 B_{\mu\nu}=0$, with
\be\label{B01}
B_{01}=-\frac{k \gamma_v}{2} \left( \frac{1}{M^2} + \frac{y_1-y_2}{M} \right) e^{-M(y_1-y_2)}\,, 
\ee 
and all unrelated components chosen to be 0. The force can be attractive or repulsive depending 
on the sign of $q_1$.

We solve equations (\ref{eomforce}) for string 1 numerically (and similarly the corresponding equations 
for string 2) starting from the simple configuration:
\be\label{initial}
 \left\{ \begin{array}{lcl} 
 {\bf x}_1 = (-\sigma \cos\alpha, - \sigma \sin\alpha, 0) \\ 
 {\bf x}_2 = (-\sigma \cos\alpha, + \sigma \sin\alpha, 0)
           \end{array}
 \right. \,,
\ee
with $\alpha=\beta_{\rm crit}$, that is, at $t=0$ we have $\beta=\beta_{\rm crit}$ in (\ref{eomforce}). 
From the above discussion we expect that the force acting locally at the junction will distort the initial 
configuration leading to a non-trivial evolution of the local angle $\beta$.  An attractive force can be 
expected to reduce the angle locally, bending the string segments from their initial straight segment 
configuration, and accelerating the junction towards the right. A repulsive force 
can be expected to have the opposite effect, increasing the angle from its equilibrium value and 
resulting in deceleration of the junction.  

Figure~\ref{fig:stringforces} shows two such examples, for $M=12$ and $k=65$. On the left plot, 
the force is attractive ($q=q_1=q_2=1$) and on the right it is repulsive ($q=-1$). The attractive force leads to 
$\beta<\beta_{\rm crit}$, as expected, bending the two strings towards each other near the junction, 
while sufficiently far away from the junction the strings retain their straight profiles at angle 
$\alpha=\beta_{\rm crit}$ from the $x$-axis. Similarly, the repulsive force produces 
$\beta>\beta_{\rm crit}$ near the junction. The former case gives rise to accelerated zipping, while for the latter, we expect unzipping to occur. Indeed, as we can see in Fig.~\ref{fig:s3repulsive}, where we have numerically solved for the evolution of $s_3(t)$, unzipping starts at $t \sim 0.25$.       

\begin{figure}[H]
\centering 
\includegraphics[scale=0.5]{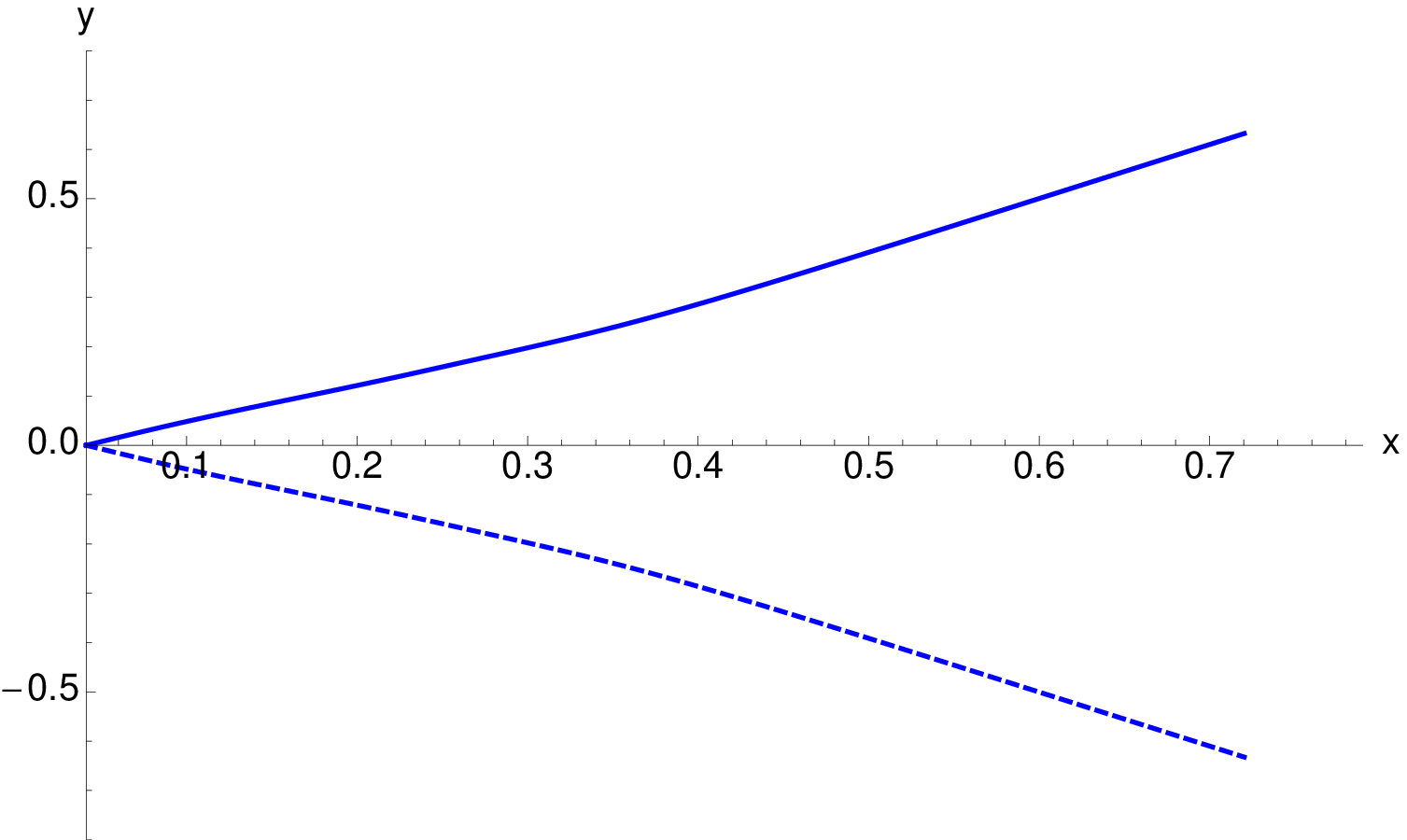}
\includegraphics[scale=0.5]{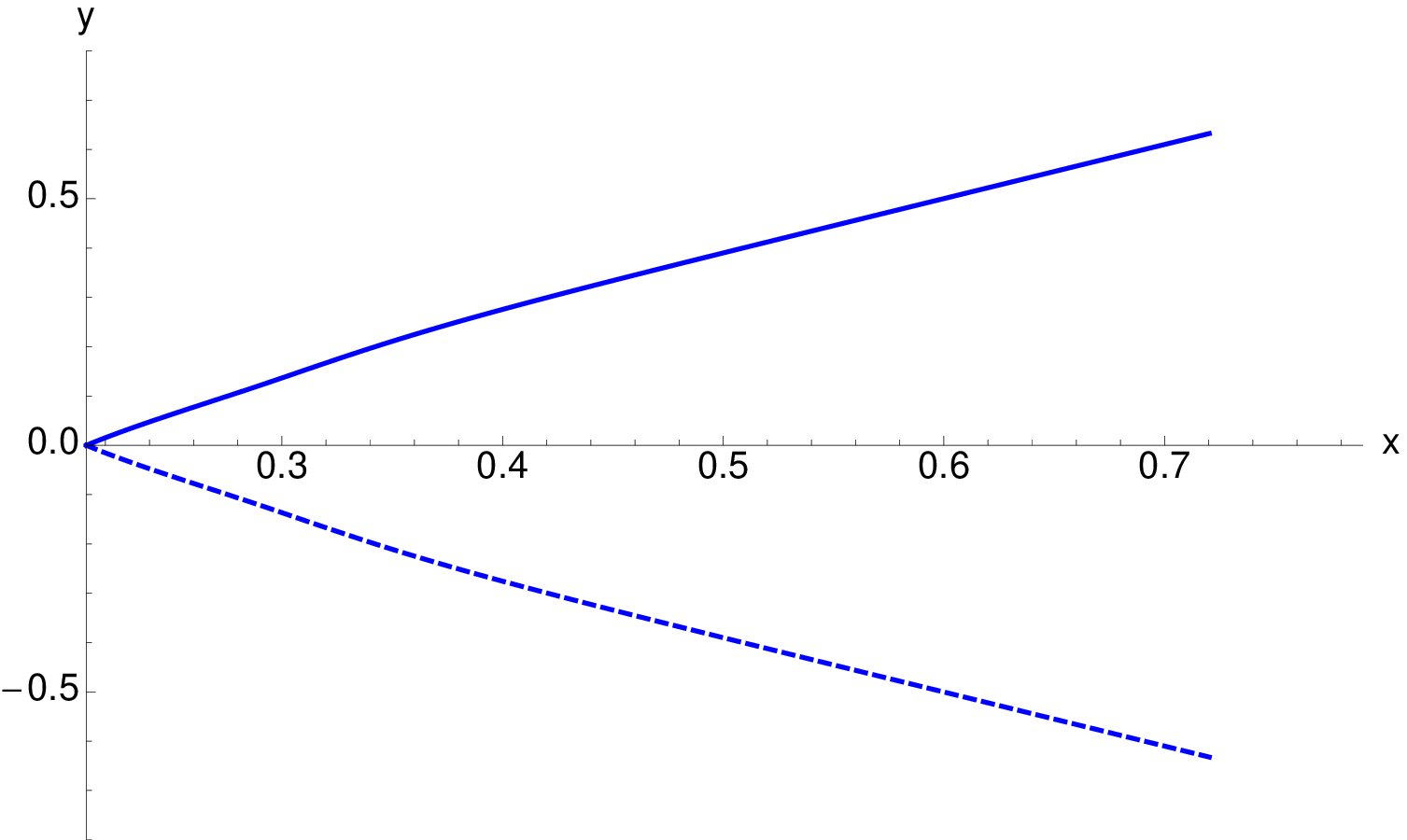}
\caption{A snapshot in the evolution of the 3-string configuration (\ref{initial}), under the influence of the 
local string force (\ref{HookYukawa}), for $M=12$ and $k=65$. On the left the force is attractive ($q=1$)
leading to $\beta<\beta_{\rm crit}$ near the junction, while on the right it is repulsive ($q=-1$) producing 
$\beta>\beta_{\rm crit}$.}
\label{fig:stringforces}
\end{figure}

\begin{figure}[H]
\centering 
\includegraphics[scale=0.6]{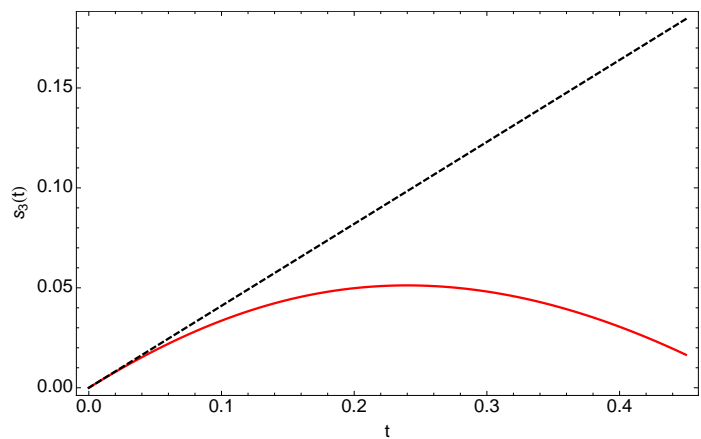}
\caption{The evolution of $s_3(t)$ under the influence of the 
local string force (\ref{HookYukawa}), for the repulsive case $M=12$ and $k=65$ (solid red line), shown together with the unperturbed case given by the CKS solution (dashed black line). Unzipping happens at $t \sim 0.25$.}
\label{fig:s3repulsive}
\end{figure}

\subsection{String Curvature -- Loops}

In this subsection we  discuss the unzipping effects of string curvature. The evolution and stability of cosmic string loops with junctions was studied in Ref.~\cite{Bevis:2009az}, using both the field theoretical and the Nambu-Goto approaches. In the field theory studies of Ref.~\cite{Bevis:2009az}, a new phenomenon occurred: the composite vortices could unzip, producing in the process new junctions whose separation could grow, destabilizing the configuration. This phenomenon was then successfully modeled within the Nambu-Goto dynamics, and the results showed that it is the initial \emph{local curvature} around a given junction that affects its evolution.

The study of junction formation and evolution following the collision of cosmic string loops has shown that unzipping
phenomena can occur
 (see Refs.~\cite{Firouzjahi:2009nt,
  Firouzjahi:2010ji}). In Ref.~\cite{Firouzjahi:2009nt} it
was shown that for colliding loops in a flat background zipping
and unzipping generically happen, while in Ref.~\cite{Firouzjahi:2010ji}
the effect of expansion of the Universe was also taken into account. In what follows we briefly review the aforementioned study and then generalise
it to include loops of unequal tensions, which is more relevant to the
case of cosmic superstring networks.

The basic picture is that of two co-planar loops extended in the
$(x,y)$-plane which are moving in the $z$-direction with opposite
velocities (Fig.~\ref{fig:loops}).  After collision, four
junctions are formed, $A, B, C$ and $D$. Certainly, due to the
symmetry of the problem, one can study the evolution of only two
junctions, for instance $B$ and $D$; $A$ and $C$ are just their mirror
images. In what follows we concentrate on junction $B$.~\footnote{The
  formalism for both junctions is essentially the same. Note
  that junction $B$ is found to unzip faster than $D$ in
  Ref.~\cite{Firouzjahi:2010ji}.}
\begin{figure}[H]
\centering 
\includegraphics[scale=0.6]{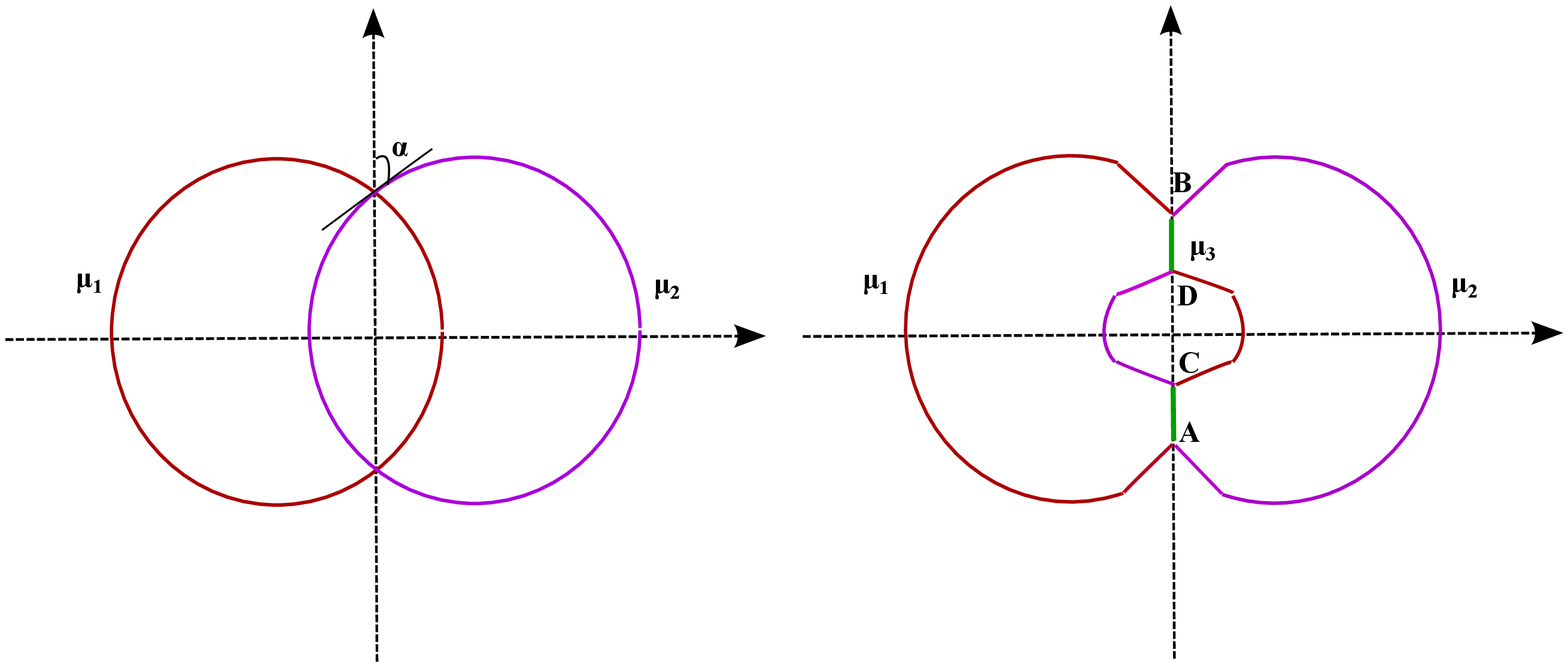}
\caption{{\it Left}: Two co-planar loops which are moving in the $z$-direction with opposite velocities collide. {\it Right}: After the collision four junctions and eight kinks are formed.}
\label{fig:loops}
\end{figure}

We consider an expanding Universe described by
Friedmann-Lema\^i}tre-Robinson-Walker (FLRW) metric
\be
\label{metric}
ds^2=a^2(\tau)(d\tau^2-d\bx^2) \, , \ee
where $\tau$ denotes the conformal time related to the cosmic time $t$
via $d t = a d \tau$ and $a(\tau)$ stands for the scale factor. We have
also assumed that the background space-time is spatially flat. The
two loops collide at $\tau=\tau_0$. 
This system is described by an action which generalizes the one of
Ref.~\cite{Copeland:2006if} to an expanding
background. Since all details can be found
in Ref.~\cite{Firouzjahi:2010ji}, here we will only present the equations
which determine the evolution of the system. Following the conventions of \cite{Firouzjahi:2010ji}, 
we denote the initial
incoming strings by $\bx_i$ with $i=1,2$, whereas the newly formed
strings are denoted by $\by_a$ with $a=1,2,3$. We also denote with
$s_i^J$ the value of the $\sigma$ coordinate of a string at a junction
$J$ (where $J$ can represent all four junctions $A,B,C,D$, but here we will concentrate on $J=B$), and with $\omega_i^J$ the value of the $\sigma$ coordinate of a
string at a kink. Note that, in general, $s$ and $\omega$ vary with
time.

The segments of old strings, which are not influenced by the
junctions, will obey the usual equations of motion in a FLRW
background, namely
\be
\frac{\partial}{\partial\tau}(\dot{\bx}_i\epsilon_{\bx_i})
+2\frac{\dot{a}}{a}\dot{\bx}_i\epsilon_{\bx_i}=
\frac{\partial}{\partial\sigma}\left(\frac{\bx'_i}{\epsilon_{\bx_i}}\right)
\, ,
\label{eomx}
\ee
where 
$\epsilon_{\bx_i}$ is defined by
$
\epsilon_{\bx_i} \equiv \sqrt{\frac{\bx_i'^2}{1- \dot \bx_i^2}}.
$
The same is true for the new strings stretched between a junction and
a nearby kink:
\be
\frac{\partial}{\partial\tau}(\dot{\by}_a\epsilon_{\by_a})
+2\frac{\dot{a}}{a}\dot{\by}_a
\epsilon_{\by_a}=
\frac{\partial}{\partial\sigma}\left(\frac{\by'_a}{\epsilon_{\by_a}}\right)~.
\label{eomy}
\ee
It can be also shown that
$
\epsilon_{\bx_i}=\epsilon_{\by_i}=\epsilon_i,
$
and 
$
\dot{\omega}_i^J\epsilon_i=-1.
$
We can define the right- and left-moving momenta $\bp_a^\pm$ as
\be
 \label{def-momenta}
  {\bp^\pm}_{\by_a}^J=\frac{\by'_a}{\epsilon_a}\pm \dot{\by}_a \, ,
  {\bp^\pm}_{\bx_i}^J=\frac{\bx'_i}{\epsilon_i}\pm \dot{\bx}_i~,
\ee
with ${\bp_a^\pm}^2=1$.  After some algebra \cite{Firouzjahi:2010ji},
we can express the unknown $\bp_{\by_a}^+$ in terms of the known
$\bp_{\by_b}^-$ and derive a system of equations for the evolution of
$\dot{s}_i^J$.  Defining
$ 
c_1 \equiv
\bp^-_{\by_2} \cdot \bp^-_{\by_3}
$
and similarly $c_2$ and $c_3$,
we obtain
\be
\label{s-eq}
1- \epsilon_1 \dot s_1^J = 
\frac{ \bar \mu M_1 (1-c_1)}{\mu_1
 \left[ M_1 (1-c_1)+ M_2 (1-c_2)+ M_3 (1-c_3) \right]}
 \, ,
\ee
where $\bar \mu=\mu_1+\mu_2+\mu_3$, $M_1 \equiv \mu_1^2 - (\mu_2 -
\mu_3)^2$ with a similar definition for $M_{2}$ and $M_3$, and $\dot
s_{2,3}$ given by the same equation with appropriate permutations of
the indices. 
Moreover, for the position of the vertex $\bY$ we find
\be
\label{Ydot-eq}
{\cal M}\dot{\bY}=-M_1(1-c_1){\bp^{-}}^J_{\by_1}-M_2(1-c_2){\bp^{-}}^J_{\by_2}
-M_3(1-c_3){\bp^{-}}^J_{\by_3}~,
\ee
with ${\cal M}=M_1 (1-c_1)+ M_2 (1-c_2)+ M_3 (1-c_3)$.
Finally, the energy conservation at the junctions implies
\be
\mu_1\epsilon_1\dot{s}^J_1+\mu_2\epsilon_2\dot{s}^J_2+\mu_3\epsilon_3\dot{s}^J_3=0~.
\ee
In Ref.~\cite{Firouzjahi:2010ji}, the previous analysis was applied to the case of two identical loops colliding. 
We are going to generalize it assuming the colliding loops have different tensions, $\mu_1 \neq \mu_2$, which is more relevant to the case of cosmic superstring networks. 
The collision happens
at $\tau=\tau_0$ and the expansion law for the scale factor is
$a(\tau)=(\tau/\tau_0)^n$, with $n=1,2$ for radiation, matter domination
respectively.  Using our conventions, with $\sigma$ increasing towards
junction $B$, we parameterize the two loops as follows (note they have the same size, for simplicity)
\bea
\bx_1=\left(b+f(\tau)\cos{\frac{\sigma_1}{R_0}},f(\tau)\sin{\frac{\sigma_1}
{R_0}},z(\tau)\right)~,
\nonumber
\\ \bx_2=\left(-b-f(\tau)\cos{\frac{\sigma_2}{R_0}},f(\tau)\sin{\frac{\sigma_2}
{R_0}},-z(\tau)\right)~,
\eea 
where $2b$ is the separation between the centres of the loops,
$R(\tau)\equiv a(\tau)f(\tau)$ is the physical radius of each loop and
$R_0=f(\tau_0)$ represents the size of each loop at the time of
collision. The independent equations of motion read
\bea
\label{back-eq1}
F'' &+& \frac{2n}{x} F' ( 1- v^2 - F'^2) + ( 1- v^2 - F'^2) F^{-1} =0~, \\
\label{back-eq2} v' &+& \frac{2n}{x} v ( 1- v^2 - F'^2) =0 \, ,
 \eea
where the loop center of mass velocity is defined by $v = \dot
z(\tau)$, and for the ease of the numerical investigations, we
introduced the dimensionless time variable $x \equiv \tau/\tau_0$ and
$F(x) \equiv f(\tau)/\tau_0$, following \cite{Firouzjahi:2010ji}. Note that prime represents derivatives
with respect to the dimensionless time $x$.

From the continuity of the left-moving momenta we find
\be {\bp^{-}}^B_{\by_1}=\left(-\sqrt{1-F'^2-v^2}\sin
\frac{\sigma_1}{R_0}-F'\cos \frac{\sigma_1}{R_0},\sqrt{1-F'^2-v^2}
\cos \frac{\sigma_1}{R_0}-F' \sin \frac{\sigma_1}{R_0},-v\right)
\nonumber \\, \ee \be {\bp^{-}}^B_{\by_2}=\left(\sqrt{1-F'^2-v^2}\sin
\frac{\sigma_2}{R_0}+F'\cos \frac{\sigma_2}{R_0},\sqrt{1-F'^2-v^2}
\cos \frac{\sigma_2}{R_0}-F' \sin \frac{\sigma_2}{R_0},v\right)~.
\nonumber \\  \ee
Since the colliding loops have unequal tensions, we first need to determine
the velocity and orientation of the joining string after collision. In
order to do that, we follow a similar treatment to the one presented
in Ref.~\cite{Copeland:2006if} for the case of straight strings colliding
in a flat background. We let string labeled 3 to lie at an angle $\theta$ to the $y$-axis,
and to move in the $z$-direction with velocity $u$ (assuming
$\mu_1>\mu_2$).  This gives
\be
   {\bp^{-}}^B_{\by_3}=(\sqrt{1-u^2}\sin{\theta},\sqrt{1-u^2}\cos{\theta},-u)~,
   \nonumber   \ee
and
\be
u^\prime +\frac{2n}{x}(1-u^2)u=0~.
\ee
The position of the vertex is \be \bY=\bx_3(s_3(\tau),\tau),
\ee which gives 
\be
\dot{\bY}=(\dot{s}_3\sin{\theta},\dot{s}_3\cos{\theta},u)~.  \ee 
At collision, the three components of the vector Eq.~(\ref{Ydot-eq}) are
\be \nonumber [{\cal
    M}\dot{s}_3+M_3(1-c_3)\sqrt{1-u^2}]\sin{\theta} |_{\tau=\tau_0}= (
\sqrt{1-F'^2-v^2}\sin a-F'\cos a )[M_1(1-c_1)-M_2(1-c_2)]|_{\tau=\tau_0}~, \ee \be \nonumber
     [{\cal M}\dot{s}_3+M_3(1-c_3)\sqrt{1-u^2}]\cos{\theta}|_{\tau=\tau_0}=
     (\sqrt{1-F'^2-v^2}\cos a + F'\sin a)[M_1(1-c_1)+M_2(1-c_2)]|_{\tau=\tau_0}~, \ee
and 
\be \nonumber
     [M_1(1-c_1)+M_2(1-c_2)]|_{\tau=\tau_0}u=[M_1(1-c_1)-M_2(1-c_2)]|_{\tau=\tau_0}v~.  \ee
After some algebra we get
\be
\frac{u_0}{v}=\frac{M_1(1-c_1)-M_2(1-c_2)}{M_1(1-c_1)+M_2(1-c_2)}|_{\tau=\tau_0}~, \ee
and 
\be \tan \theta=\left(\frac{u_0}{v}\right)\left(\frac{ \sqrt{1-F'^2-v^2}\sin
  a-F'\cos a}{\sqrt{1-F'^2-v^2}\cos a + F'\sin
  a}\right)|_{\tau=\tau_0}~.  \ee
As a first check, solving the above system of equations for
$\mu_1=\mu_2$, we find $u_0=\theta=0$, as expected, and solving the derived
equations numerically we reproduce the results obtained in
Ref.~\cite{Firouzjahi:2010ji}.

Now let us try an asymmetric example, assuming we are in the radiation dominated era. As initial conditions we choose $a=\pi/9$, $\mu_1=2$,
$\mu_2=1$, $\mu_3=2.5$, $v(x=1)=0.4$, and $F'(x=1)=0.1$, and we define $S_i \equiv s_i/R_0$ \cite{Firouzjahi:2010ji} to simplify the equations and the numerical analysis. We indeed find two real and
positive solutions, which are $u(x=1) \rightarrow 0.168$ and $\theta
\rightarrow 0.102$.  Now we can proceed to solve for the evolution of
the junction (with $F(x=1)=100$). We see that $S_3$ initially increases but after some time the growth of the zipper stops
and unzipping occurs --- note that this happens before the loops shrink to zero. 

\begin{figure}[H]
\begin{center}
\includegraphics[width=3in,height=2in]{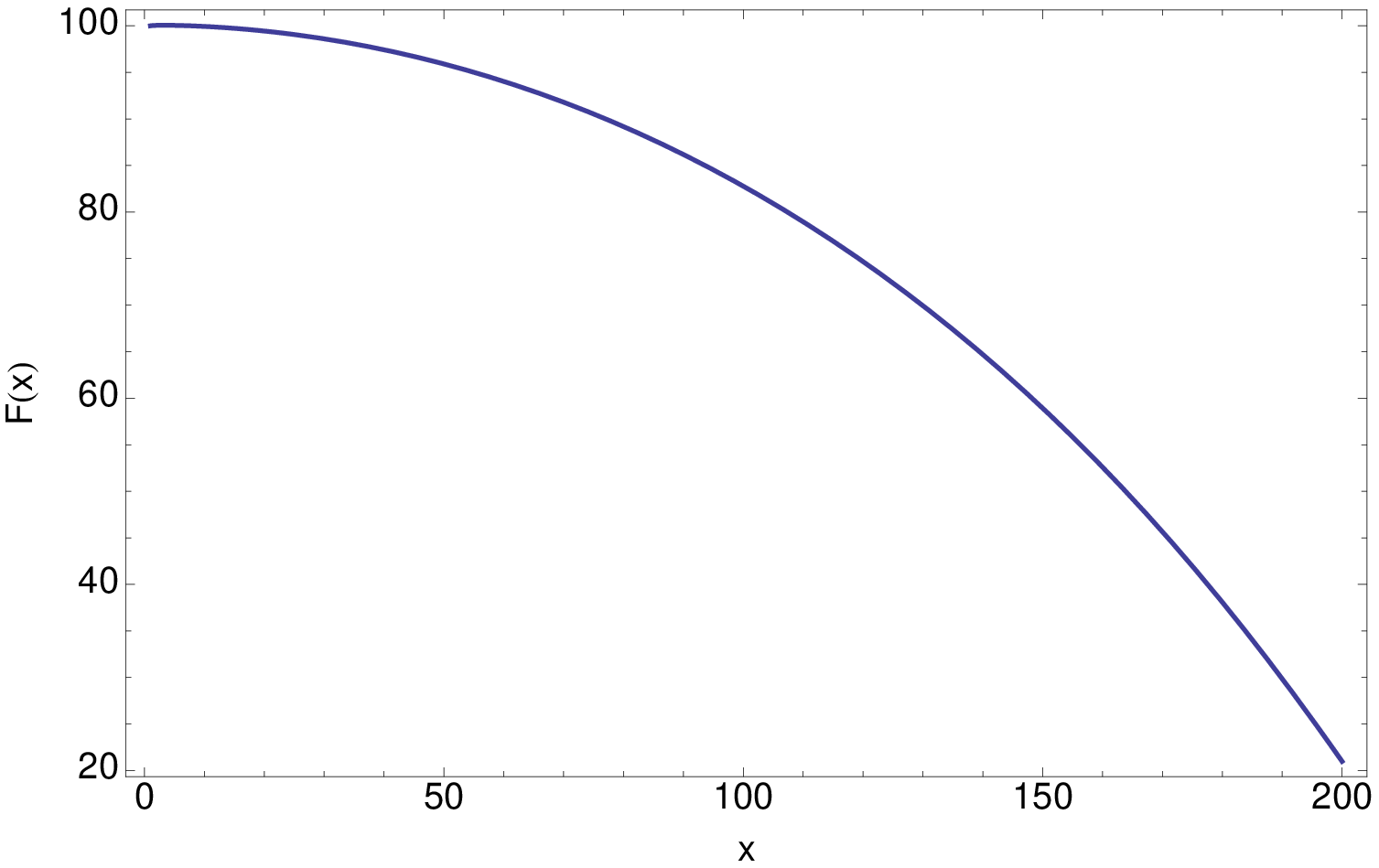}
\includegraphics[width=3in,height=2in]{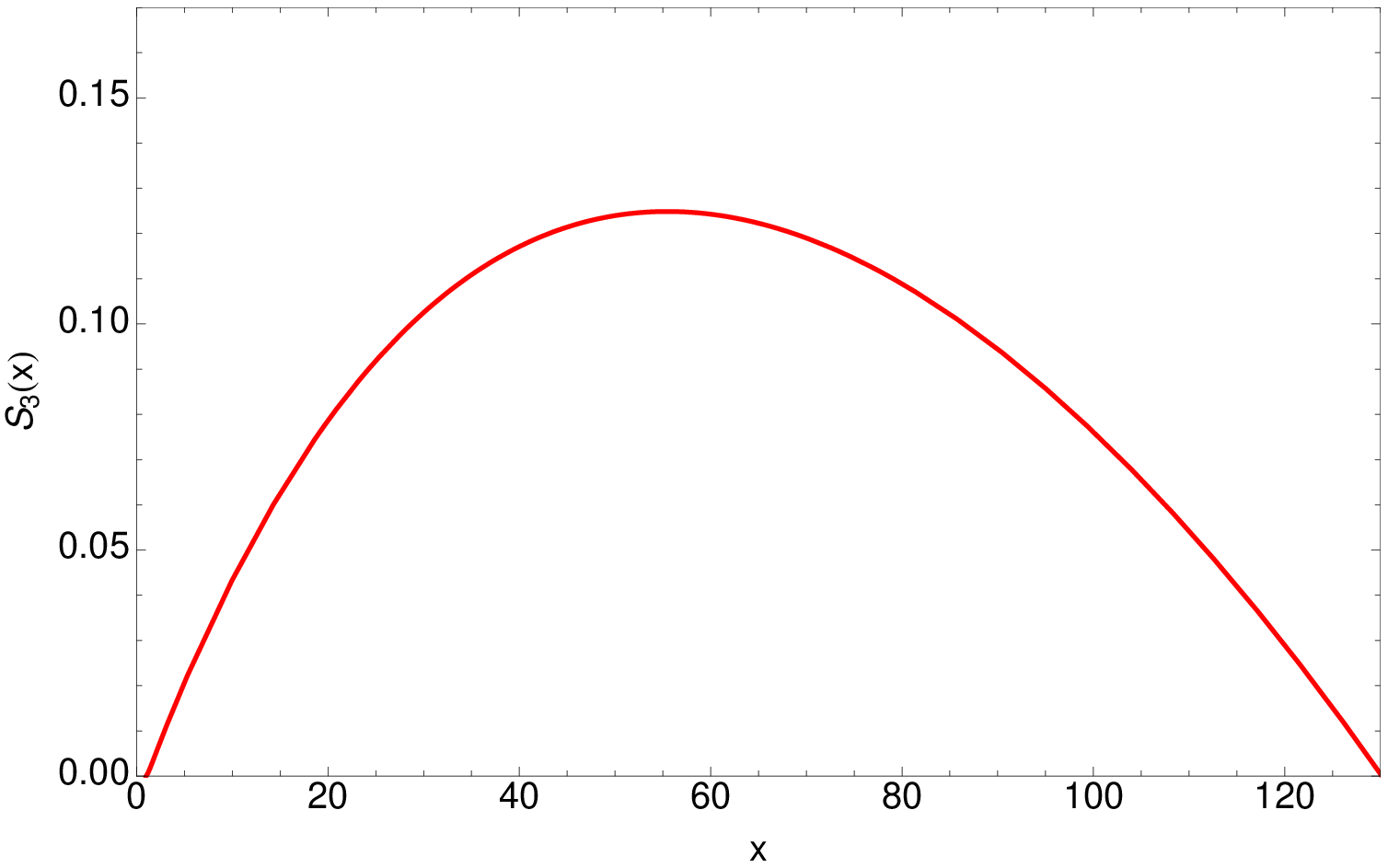}
\caption{\textit{Left panel}: The background evolution for F(x) with
  $F(x=1)=100$, $F'(x=1)=0.1$, and $v(x=1)=0.4$ for the radiation
  era. \textit{Right panel}: The evolution of junction B for the
  radiation dominated background, for $a=\pi/9$, $\mu_1=2$, $\mu_2=1$,
  $\mu_3=2.5$, $v(x=1)=0.4$, $u(x=1)=0.168$, $\theta=0.102$.}
\label{case2}
\end{center}
\end{figure}

As in Ref.~\cite{Firouzjahi:2010ji}, the initial condition $F(x=1)$ is a measure of the physical radius of 
the colliding loops compared to the Hubble radius. In our case, where $F(x=1)=100$, we considered loops of superhorizon size. The qualitative results of our study are the same as the ones of Ref.~\cite{Firouzjahi:2010ji}: As we know, large superhorizon loops can be approximated with straight strings. Consequently, if the initial conditions are appropriate for junction formation the junction will grow, following the usual behavior of straight infinite strings. However, at some point these loops will reenter the horizon, their velocities will rapidly reduce and unzipping will occur. In our example (Fig.~\ref{case2}), 
after junction formation $s^B_3$ reaches a maximum value indicating its unzipping. Similarly, junction $D$ also unzips, although the unzipping of junction $B$ happens sooner. When junctions $B$ and $D$ meet, the loops disentangle. However, there is also the possibility that the loops shrink to zero before $B$ and $D$ meet. For intermediate and small size loops the results are similar, but the smaller the loop is the less they depend on the background expansion, as expected. In general, we can conclude that string curvature and loop dynamics are an effective unzipping mechanism.

\section{Conclusions}
\label{sec:concls}
In this paper we have studied the dynamics of string junctions in an attempt to identify dynamical mechanisms that could trigger the unzipping of Y-junction configurations. We are motivated by field theory simulations of string networks with junctions~\cite{Rajantie:2007hp,Urrestilla:2007yw}, where one observes a lower than expected abundance of heavy string segments. This may suggest that junction formation may be dynamically obstructed or that there is a tendency for heavy string zippers to unzip after they form. One could entertain the possibility that a Y-junction formed by the collision of two string segments could unzip for high enough collision velocities, as the colliding strings retain their original motion beyond the kinks, effectively pulling the zipper apart. However, a simple exercise shows that this can only work for non-relativistic strings as it requires the collision velocity to be larger than the speed of propagation of kinks along the string. For relativistic strings, having worldsheet Lorentz invariance, kinks propagate at the speed of light and this kinematic mechanism cannot work. A more relevant physical effect for string networks is perhaps velocity damping due to cosmic expansion, but this can only slow down (not invert) the zipping process and is also suppressed at sub-horizon scales. 

Having in mind cosmic superstring models that are generally expected to form junctions, we have in this paper investigated a number of potential dynamical mechanisms that could be responsible for unzipping in string networks. Firstly, we have studied, within the Nambu-Goto approximation, the stability of massive string junctions under the influence of the tensions of three strings joining in a Y-type configuration, and concluded that these configurations are stable under deformations of the tension balance condition at the junction. This justifies the usual assumption of string evolution models with junctions that zippers grow according to the special solutions of \cite{Copeland:2006eh,Copeland:2006if,Copeland:2007nv} once formed. Secondly, we have investigated whether monopole or string forces, and string curvature for loops with junctions, can lead to unzipping of string junctions.  In each case we have found solutions exhibiting decelerating zipping leading to the unzipping of Y-type configurations, and we have discussed the conditions under which unzipping happens in the context of our simple 3-string Nambu-Goto modelling. 

It remains unclear at this stage whether these mechanisms can play a significant r\^ole in realistic string networks. If such an unzipping mechanism gets realised within cosmic (super)string networks it would affect the relative abundance between the two lightest string species, which, as was shown in \cite{PACPS,ACMPPS}, controls the characterstically stringy B-mode signal of these networks.  To date, the incorporation of realistic string interactions in the modelling of these networks has not been achieved and our results provide further motivation for this.  

Note that the Nambu-Goto equations for a massive junction, which we have obtained, do not have a well-defined massless ($m \rightarrow 0$) limit. This has been already discussed in Ref.~\cite{Siemens:2000ty} for a system of (anti)monopoles connecting strings, which are expected to eventually collide and annihilate being constrained to live on a string. A similar situation may be expected in the case of a massive junction after unzipping takes places (i.e. when a massive junction meets its mirror).

\section*{Acknowledgements}
The work of A.A. was supported by Marie Curie grant FP7-PEOPLE-2010-IEF-274326 at the University of Nottingham,
and by a Nottingham Research Fellowship. A.\ P.\ acknowledges support from STFC grant ST/H002774/1. A.P. also 
received support by the project GLENCO, funded under the FP7, Ideas, Grant Agreement n. 259349, while part of this 
work was in progress at the University of Bologna. A.\ P.\ acknowledges the University of Nottingham for hospitality during 
various stages of this work.

\end{document}